\documentstyle[aps,epsf]{revtex}

\begin{document}

\author{Xiao-jun Li and M. Schick\\
        Department of Physics, Box 351560 \\
        University of Washington, Seattle 98195-1560}
\title{Theory of Lipid Polymorphism: Application to Phosphatidylethanolamine
and  Phosphatidylserine}
\date{\today}

\maketitle

\begin{abstract}
  We introduce a microscopic model of a lipid with a charged headgroup
and flexible hydrophobic tails, a neutral solvent, and counter ions.
Short-ranged interactions between hydrophilic and hydrophobic moieties
are included as are the Coulomb interactions between charges. Further,
we include a short-ranged interaction between charges and neutral
solvent, which mimics the short-ranged, thermally averaged interaction
between charges and water dipoles.  We show that the model of the
uncharged lipid displays the usual lyotropic phases as a function of the
relative volume fraction of the headgroup. Choosing model parameters
appropriate to dioleoylphosphatidylethanolamine in water, we obtain 
phase behavior which agrees well with experiment. Finally we
choose a solvent concentration and temperature at which the uncharged
lipid exhibits an inverted hexagonal phase and turn on the headgroup
charge.  The lipid system makes a transition from the inverted hexagonal
to the lamellar phase which is related to the increased waters of
hydration correlated with the increased headgroup charge via the
charge-solvent interaction.  The polymorphism displayed upon variation
of pH mimics that of the behavior of phosphatidylserine.
\end{abstract}

\section{Introduction}

Biological lipids in solution display several different lyotropic
phases, and the implications this may have for biological function has been a
subject of speculation for many years (Cullis et al., 1985; de Kruijff
1997). Lipid phase
behavior depends upon several factors, some of which are
intrinsic to the lipid architecture itself. 
For example, an increase in the length of the hydrocarbon tails brings about
transitions from lamellar, $L_{\alpha}$, to inverted hexagonal,
$H_{II}$, phases (Seddon, 1990), while an increase in the volume of the
headgroup brings about the reverse (Gruner, 1989). Other factors
regulating phase behavior are externally controlled, such as 
temperature, solvent concentration, and 
solvent pH (Hope and Cullis, 1980, Seddon et al., 1983, 
Bezrukov et al., 1999). 
It is these factors which are the focus of this paper. 

Lipid phase behavior has been addressed extensively by the construction
of phenomenological free energy functions which contain terms
describing, {\em inter alia}, bending, hydration, and interstitial
energies (Helfrich, 1973; Kirk et al., 1984; Rand and Parsegian, 1989;
Kozlov et al., 1994).  Such approaches, which obtain their several
parameters from experimental measurement of various quantities, are
quite useful, particularly in correlating phase behavior with other
thermodynamic properties. Nonetheless, it would clearly be desirable to
derive all thermodynamic quantities, including the phase
behavior, by applying statistical mechanics to 
a microscopic model of the system. In addition
to simplifying the description considerably, such approaches would
correlate phase behavior with the the architectural properties of the
lipid itself and its solvent.

Analytic,
mean-field, approaches of statistical mechanics 
have been applied to anhydrous lipids to investigate
behavior of increasing complexity.  Such methods have been combined with
realistic models of lipid tails to determine how the hydrocarbon chains
pack in aggregates and in bilayers (Marcelja, 1974; Gruen, 1981 and
1985; Ben-Shaul et al. 1985;
Fattal and Ben-Shaul, 1994). Results for the bilayer are in good agreement with
molecular dynamic simulation (Heller et al., 1993). 
These methods have shown that in a
neutral, anhydrous system, the entropy of the lipid tails always favors
the $H_{II}$ over the $L_{\alpha}$ phase, and that a change in area per
headgroup could bring about a transition between them (Steenhuizen et
al., 1991).

Aggregates, such as the lipid bilayer, in the presence of solvent have
also been considered within the mean-field approach applied to lattice
models (Leermakers and Scheutjens, 1988). In addition to the tails, one
must now model the solvent and the headgroups, and   phosphatidylcholine and
phosphatidylserine headgroups are among those which have been described
(Meijer et al., 1994).  The method is flexible and has been applied to
many different systems, including bilayers with trans-membrane guest
molecules (Leermakers et al., 1990). Results are quite good, with the
exception that the local volume fraction of solvent inside the bilayer
is rather large, several orders of magnitude greater than that observed
in experiment (Jacobs and White, 1989). Lattice models, however, are
not well-suited to the description of transitions between phases of different
symmetry.

Recently  
similar methods were applied to a system of
solvent and monoacyl lipid embedded in a continuous space, and  
the phase diagram was obtained by solving the mean-field
theory exactly (M{\"u}ller and Schick, 1998). It
displayed both $L_{\alpha}$ and $H_{II}$ phases, so that  the
transition between them could be studied as a function of lipid architecture. 
The dependence of the transition
on the architectural parameters, length of tail and volume of headgroup,
was that observed in experiment. However, the fraction of solvent within
the bilayers was again too large. 

In this paper, we introduce a computationally more tractable model of a
lipid than that employed by M{\"u}ller and Schick, one whose hydrocarbon
tails are modeled as flexible chains rather than within the rotational
isomeric states framework (Flory, 1969; Mattice and Suter, 1994) 
employed earlier. We first
study the model with an uncharged headgroup. Its phase
behavior, both with respect to variations in architecture and variations
in solvent concentration, is as expected, and in agreement with
experiment. In particular,  choosing model parameters appropriate to
dioleoylphosphatidylethanolamine (DOPE), we obtain a phase diagram 
similar to that observed (Gawrisch et al. 1992, Kozlov et al. 1994). We 
extract the variation with temperature and solvent concentration 
of the lattice parameter of the inverted hexagonal phase, and compare it to
experiment (Tate and Gruner, 1989, Rand and Fuller, 1994). 
The agreement is excellent. We also find that the concentration of 
solvent within the bilayer is vanishingly small. 
 We then  allow the headgroup to be negatively charged,
introduce counter ions into the system, and include the Coulomb
interaction between all charges. We show that the Coulomb repulsion
between the headgroups does {\em not} tend to stabilize the $L_{\alpha}$
phase at the expense of the $H_{II}$, but has the {\em opposite}
effect. Lastly we turn on a short-ranged interaction between charges and
neutral solvent, an interaction which models the thermally averaged
interaction between charges and the dipole of water. We then find that as
the charge on the headgroup is turned on, the $L_{\alpha}$ phase {\em
is} stabilized with respect to the $H_{II}$. In effect, as the charge on
the headgroup increases, so too do the waters of hydration. In addition, the
counter ions which are attracted to the headgroup  are also enlarged by their
own waters of hydration. It is the totality of these
waters which effectively increases the headgroup volume and therefore
stabilizes the lamellar phase.

The paper is organized as follows. In the next section, we introduce the
model for the charged lipid, the solvent, and counter ions, specify all
the interactions between them and set up the partition function of the
system. In section III, we derive the self-consistent field theory for
it. At the heart of the theory are four self-consistent equations for
the electrostatic potential of the system and the three effective fields
which determine the headgroup, tail, and solvent densities.  One of
these self-consistent conditions is simply the non-linear
Poisson-Boltzmann equation. In section IV, we expand all functions of
position into a complete set of functions having a specified space-group
symmetry, and rewrite the self-consistent
equations in terms of the coefficients of these expansions. These
equations are solved numerically, and the free energies of the various
phases computed. A comparison of the free energies yields the phase
diagram. In Section V, we present the phase diagram for the neutral
lipid as a function of temperature and one architectural parameter. We
include here only the classical phases, lamellar, inverted and normal
hexagonal, and inverted and normal body-centered-cubic, as well as the
disordered phase. For the remainder of the section, we choose 
an architecture such that the anhydrous, neutral lipid
orders into  the $H_{II}$ phase. Results for the system in
the presence of a neutral solvent, along with comparisons 
to experiment, are presented next.
 
In Section VI, we consider the charged lipid. We choose a 
water concentration such that the
neutral lipid remains in the $H_{II}$ phase. By varying the counter ion
concentration, we turn on the charge on the headgroup, and thus all
Coulomb interactions, but we keep the short-ranged interaction between
charges and neutral solvent set to zero. We show that the $H_{II}$ phase
is stabilized with respect to the $L_{\alpha}$ rather than destabilized,
a result completely analogous to that of polyelectrolytes (Nyrkova et
al., 1994).  Lastly we present our results for the case in which, in
addition to the Coulomb interaction, a short-ranged interaction between
charges and solvent is turned on.  Here we find that the $L_{\alpha}$ is
indeed stabilized with respect to the $H_{II}$ phase, in agreement with
experiment (Hope and Cullis, 1980, Bezrukov et al., 1999).  

\section{The Model}
We consider a system composed of charged lipids, neutral solvent, and
counter ions in a volume $V$. There are $n_L$ lipids, each of which consists 
of a head, with volume
$v_h$, and two  equal length, completely flexible,  
tails each consisting of $N$ segments of 
volume $v_t$. Each lipid tail is characterized by a  
radius of gyration
$R_g=(Na^2/6)^{1/2}$, with $a$ the statistical segment length.
The heads carry a negative charge
$-eQ_h$. The solvent consists of $n_s$ neutral particles of volume $v_s$,
while the $n_c$ counter ions have charge $+e$ and negligible volume,
$v_c=0$. There are five dimensionless densities which totally specify
the state of the system; the number density of 
the headgroups, $\hat\Phi_h$, of the
tail segments, $\hat\Phi_t$, and of the solvent, $\hat\Phi_s$, and the
charge density of the headgroups, $e\hat P_h$, and of the counter ions,
$e\hat P_c$. They can be written as
\begin{eqnarray}
\label{head}
\hat\Phi_h({\bf r})&=&v_h\sum_{l=1}^{n_l}\delta({\bf r}-{\bf r}_l(1/2)),\\
\hat\Phi_t({\bf r})&=&v_h\sum_{l=1}^{n_l}\int_0^1 
\delta({\bf  r}-{\bf r}_l(s))ds, \\
\hat\Phi_s({\bf r})&=&v_h\sum_{j=1}^{n_s}\delta({\bf r}-{\bf R}_{s,j}),\\
\hat P_h({\bf r})&=&-v_h\sum_{l=1}^{n_l}Q_{h,l}\delta({\bf r}-{\bf
  r}_l(1/2)),\\
\hat P_c({\bf r})&=&v_h\sum_{i=1}^{n_c}\delta({\bf r}-{\bf R}_{c,i}).
\end{eqnarray}
We have chosen $v_h$ as a convenient volume to make all densities 
dimensionless. In the above,
 ${\bf R}_{s,j}$ is the position of the $j$'th solvent particle, and 
${\bf R}_{c,i}$ the position of the $i$'th counter ion. The
configuration 
of the $l$'th lipid is described by a space curve ${\bf r}_l(s)$,
where $s$ ranges from 0,  at the end of one tail, through $s=1/2$ at
which the head is located, to $s=1$,
the end of the other tail. The nominal probability that the charge 
on the headgroup of the 
$l$'th lipid, 
$-e\ Q_{h,l}$, is equal to $-e$ or 0 is $p$ or $1-p$
respectively. As we model the case in which charges can associate or
disassociate from the headgroup, it will be necessary to average the
partition function of the system with respect to the charge
distribution. 
This corresponds to an annealed distribution in the nomenclature of 
Borukhov (Borukhov et al., 1998). The concentrations of lipid, 
solvent, and free counter ions are controlled by chemical potentials. In 
particular, increasing the number of {\em free}, positive, counter ions
 implies, by charge 
neutrality, an increase in the negative charge on the headgroups, and 
thus corresponds to an increase in the pH of the system. 

The interactions among these elements are as follows. First there is a
repulsive, contact interaction between headgroup
and tail segments, and also between solvent and tail segments. The strength
of the interaction is  $kTv_h\chi$, where $k$ is Boltzmann's constant and $T$
the absolute
temperature. 
Second there is the  Coulomb interaction between all charges. The
dielectric constant of the solvent is denoted $\epsilon$. Finally
there is a contact interaction between all charges and the neutral
solvent whose strength  is  $kTv_h \lambda$. 
This is to model the
short-ranged, thermally averaged,  interaction between charges and the
dipole of water, an attractive interaction which decreases like $r^{-4}$
and is of
strength $e^2u^2/6\epsilon^2kT$, where $u$ is the dipole moment of water
(Israelachvili, 1985).
Thus the energy per unit volume of the system, $E/V$,  
can be written
\begin{eqnarray}
\label{energy}
{v_h\over kT}{E\over V}[\hat\Phi_h,\hat\Phi_t,\hat\Phi_s,\hat P_h,\hat P_c]&=&
2\chi N\int{d{\bf r} \over V}[\hat\Phi_h({\bf r})+\hat\Phi_s({\bf r})]
\hat\Phi_t({\bf r})\nonumber \\
& &+{\beta^*\over 8\pi}\int {d{\bf r}\over
  V}{d{\bf r'}\over R_g^2}
  [\hat P_h({\bf r})+\hat P_c({\bf r})]{1\over |{\bf r}-{\bf r'}|}
[\hat P_h({\bf r'})+\hat P_c({\bf r'})]\nonumber \\
& & -\lambda\int
{d{\bf r}\over V}\hat\Phi_s({\bf r})[\hat P_c({\bf r})-\hat P_h({\bf
  r})],
\end{eqnarray}
where 
\begin{equation}
\label{betastar}
\beta^*\equiv{4\pi e^2R_g^2\over
v_h\epsilon kT}
\end{equation}
is a dimensionless measure of the strength of the Coulomb interaction.  
The grand partition function (Matsen, 1995) of the system is
\begin{eqnarray}
\label{pf}
{\cal Z}&=&\sum_{n_l,n_c,n_s}{z_l^{n_l}z_c^{n_c}z_s^{n_s}\over
  n_l!n_c!n_s!}\int\prod_{l=1}^{n_l}\tilde{\cal
    D}{\bf r}_l\tilde{\cal D}Q_{h,l}
\prod_{i=1}^{n_c}d{\bf R}_{c,i}\prod_{j=1}^{n_s}d{\bf
R}_{s,j}\nonumber \\
&\times&
\exp\left\{ {-E[\hat\Phi_h,\hat\Phi_t,\hat\Phi_s,\hat P_h,\hat P_c]/kT}\right\}
\delta(1-\hat\Phi_h-\gamma_s\hat\Phi_s-\gamma_t\hat\Phi_t).
\end{eqnarray}
Here $\int\tilde{\cal D}{\bf r}_l$ denotes a functional integral over the
possible configurations of the $l$'th lipid and in which, in
addition to the Boltzmann weight, the
path is weighted by the  
factor ${\cal P}[{\bf r}_{t,l}(s);0,1],$ with
\begin{equation}
{\cal P}[{\bf r},s_1,s_2]={\cal N}\exp\left[-{1\over 8R_g^2}
\int_{s_1}^{s_2}ds |{d{\bf
    r}(s)\over ds}|^2\right],
\end{equation}
with ${\cal N}$ an
unimportant normalization constant. The notation $\int\tilde{\cal
  D}Q_{h,l}$ denotes an integral over the probability distribution of
the charge on the headgroup of the $l$'th lipid. 
We have enforced an incompressibility constraint on
the system
with the aid of the delta function 
$\delta(1-\hat\Phi_h-\gamma_s\hat\Phi_s-\gamma_t\hat\Phi_t)$, where
$\gamma_s=v_s/v_h$, and $\gamma_t=2Nv_t/v_h$. The latter parameter is the
lipid architectural parameter. The relative volume of the headgroup
with respect to that of the entire molecule is $1/(1+\gamma_t)$.

The model is now completely defined. The solvent is specified by
$\gamma_s$, its
volume per particle relative to that of the headgroup, 
and the architecture of the lipid
is characterized by $\gamma_t$. 
There are three interactions, hydrophobic-hydrophilic,
charge-charge, and charge-solvent, whose strengths are given by $\chi$,
$\beta^*$, and $\lambda$ respectively. The external parameters are the
temperature, conveniently specified in terms of a dimensionless temperature 
$T^*\equiv (2\chi N)^{-1}$, the
fugacity of the solvent, $z_s$, and the fugacity of the free
counter ions, $z_c$, which, by charge neutrality, controls the charge on
the lipid headgroups. The characteristic length in the system is the
radius of gyration, $R_g$.
In the next two sections we derive 
the self consistent field theory for the model, first in real space,
 and 
then in the following section, in Fourier space.
\section{Theory: real space}
Evaluation of the partition function of Eq. \ref{pf} is difficult
because the
interactions are products of densities, each of which depends on the
{\em specific} coordinates of one of the elements of the system. This 
dependence is eliminated in a standard way. We illustrate it on
$\hat\Phi_h({\bf r})$ which, from its definition in Eq. \ref{head}, depends
on the coordinates of the headgroup, ${\bf r}_l(1/2)$. One introduces into the
partition function the identity
\begin{eqnarray}
1&=&\int{\cal D}\Phi_h\delta(\Phi_h-\hat\Phi_h),\nonumber\\
 &=&\int{\cal D}\Phi_h{\cal D}W_h\exp\left\{ {1\over v_h}\int W_h({\bf r})
[\Phi_h({\bf r})-\hat\Phi_h({\bf r})]d{\bf r}\right\},
\end{eqnarray}
in which  
$\Phi_h({\bf r})$ does {\em not} depend on any specific coordinates of one 
of the
elements of the system, but is simply a function of ${\bf r}$.
The
integration on $W_h$ extends up the imaginary axis. Inserting such
identities 
for the five densities $\hat\Phi_h$, $\hat\Phi_t$, $\hat\Phi_s$,
$\hat P_h$,
and $\hat P_c$, and a similar identity for the delta function expressing
the incompressibility
condition, one rewrites the partition function, Eq. \ref{pf}, as 
\begin{eqnarray}
{\cal Z}&=&
\label{pf2}
\int{\cal
    D}\Phi_h{\cal D}W_h{\cal D}\Phi_t{\cal D}W_t{\cal D}\Phi_s{\cal
    D}W_s{\cal D} P_h{\cal D}U_h{\cal D} P_c{\cal D}U_c{\cal
    D}\Xi\nonumber \\
   &\times& \exp\{z_l{\cal Q}_l[W_h,W_t,U_h]+z_c{\cal Q}_c[U_c]+
z_s{\cal Q}_s[W_s] -E[\Phi_h,\Phi_t,\Phi_s, P_h, P_c]/kT \}
    \nonumber \\
 &\times& \exp\left\{ {1\over v_h}\int 
\left[W_h\Phi_h+W_t\Phi_t+W_s\Phi_s+U_h P_h+U_c P_c+
\Xi(1-\Phi_h-\gamma_s\Phi_s-\gamma_t\Phi_t)\right]d{\bf r}\right\},
\end{eqnarray}
where 
\begin{equation}
\label{lipidpf1}
{\cal Q}_l[W_h,W_t,U_h]=\int \tilde{\cal
    D}{\bf r}_l\tilde{\cal D}Q_{h}
\exp\left\{-W_h({\bf r}_l(1/2))+Q_{h}U_h({\bf r}_l(1/2))-
\int_0^1ds W_t({\bf  r}_l(s))\right\},
\end{equation}
is the partition function of a single lipid in external fields $W_h$,
$W_t$, and $U_h$,
\begin{equation}
\label{pfc}
{\cal Q}_c[U_c]=\int d{\bf R}_{c}\exp[-U_c({\bf R}_{c})],
\end{equation}
is the partition function of a single counter ion of unit positive
charge in an external potential
$U_c$, and
\begin{equation}
\label{pfs}
{\cal Q}_s[W_s]=\int d{\bf R}_s\exp[-W_s({\bf R}_s)],
\end{equation}
is the partition function of a single solvent particle in the external
field $W_s$.
It is convenient to shift the zero of all chemical potentials so that 
$z_l\rightarrow 1/v_h$, $z_c\rightarrow z_c/v_h$, and $z_s\rightarrow
z_s/v_h$. The partition function, Eq. \ref{pf2}, can then be 
written in the form
\begin{equation}
\label{pf3}
{\cal Z}=\int{\cal
    D}\Phi_h{\cal D}W_h{\cal D}\Phi_t{\cal D}W_t{\cal D}\Phi_s{\cal
    D}W_s{\cal D} P_h{\cal D}U_h{\cal D} P_c{\cal D}U_c{\cal 
    D}\Xi\exp[-\Omega/kT],
\end{equation}
with
\begin{eqnarray}
\label{free1}
{v_h\over kTV}{\Omega}
& &=-{{\cal Q}_l[W_h,W_t,U_h]\over V}-
z_c{{\cal Q}_c[U_c]\over V}
-z_s{{\cal Q}_s[W_s]\over V}+
{v_h\over kT V}E[\Phi_h,\Phi_t,\Phi_s,P_h,P_c]\nonumber \\
&-&
\int{d{\bf r}\over V}[W_h\Phi_h+W_t\Phi_t+W_s\Phi_s+U_hP_h+U_cP_c
+\Xi(1-\Phi_h-\gamma_s\Phi_s-\gamma_t\Phi_t)].
\end{eqnarray}
No approximations have been made to this point. What has been
accomplished is a rewriting of the partition function from a form, Eq
\ref{pf}, in
which all entities interact directly with one another, to a form, 
Eqs. \ref{pf3} and \ref{free1}, in
which they interact indirectly with one another via fluctuating fields.
Although the integrals in Eq. \ref{pf3} over $\Phi_h,\ \Phi_t,\ \Phi_s,\ P_h,
\ P_c$ and $\Xi$ could all be carried out, as they are no worse than Gaussian,
the integrals over the fields $W_h,\ W_t,\ W_s,\ U_h,$ and $U_c$
cannot. Therefore we employ the self-consistent field theory in
which we replace the integral in Eq. \ref{pf3} by its integrand
evaluated at its extremum. The values of $W_h$, $\Phi_h$, etc. 
which satisfy the extremum
conditions will be denoted by the corresponding lower case letters
$w_h$, and $\phi_h$, etc. The equations which determine them are six 
self-consistent equations for the six fields $w_h$, $w_t$, $w_s$, $u_h$, 
$u_c$, and $\xi$. They are
\begin{eqnarray}
\label{sc1}
w_h({\bf r})&=&2\chi N\phi_t({\bf r})+ \xi({\bf r})\\
w_t({\bf r})&=&2\chi N(\phi_h({\bf r})+\phi_s({\bf r}))+\gamma_t\xi({\bf r})\\
w_s({\bf r})&=&2\chi N\phi_t({\bf r})-\lambda(\rho_c({\bf
  r})-\rho_h({\bf r}))+\gamma_s\xi({\bf r})\\
\label{pb}
u({\bf r})&\equiv&{u_h({\bf r})+u_c({\bf r})}\over 2\nonumber\\
 &=& {\beta^*\over 4\pi}\int {d{\bf r}'\over R_g^2}{\rho_h({\bf
     r}')+\rho_c({\bf r}')
\over|{\bf r}-{\bf r}'|}\\
\label{us}
u_s({\bf r})&\equiv&{u_h({\bf r})-u_c({\bf r})\over 2}\nonumber\\
            &=&\lambda\phi_s({\bf r})\\
\label{incompressibility}
1&=&\phi_h({\bf r})+\gamma_t\phi_t({\bf r})+\gamma_s\phi_s({\bf r}).
\end{eqnarray}
As the field $\xi$ is easily eliminated, the six equations readily reduce 
to five. The simplicity of Eq. \ref{us} reduces this, in practice, to a
set of four equations.
The five densities $\phi_h$, $\phi_t$, $\phi_s$, $\rho_h$, and $\rho_c$ 
are functionals of all of the above fields except  $\xi$, and 
therefore close the 
cycle of self-consistent equations:
\begin{eqnarray}            
\label{phih}
\phi_h({\bf r})[w_h,w_t,u_h]&=&-{\delta{\cal Q}_l[w_h,w_t,u_h]\over\delta w_h({\bf r})}\\
\label{phit}
\phi_t({\bf r})[w_h,w_t,u_h]&=&-{\delta{\cal Q}_l[w_h,w_t,u_h]\over\delta w_t({\bf r})}\\
\phi_s({\bf r})[w_s]&=&-z_s{\delta{\cal Q}_s[w_s]\over\delta w_s({\bf r})}\\
\label{phis}
               &=&z_s\exp[-w_s({\bf r})]\\
\label{headcharge}
\rho_h({\bf r})[w_h,w_t,u_h]&\equiv&-{\delta{\cal Q}_l[w_h,w_t,u_h]\over\delta u_h({\bf r})}\\
\label{rhoc}
\rho_c({\bf r})[u_c]&=&-{\delta{\cal Q}_c[u_c]\over\delta u_c({\bf
    r})}\\
\label{rhoc2}
               &=&z_c\exp[-u_c({\bf r})]
\end{eqnarray}
The density $\phi_h({\bf r})$ is simply the expectation value of
$\hat\Phi_h({\bf r})$ in the single lipid ensemble. Similar interpretations
follow for the other densities. 
Note that one of the self-consistent equations, 
Eq. \ref{pb}, is simply the non-linear Poisson-Boltzmann equation, and 
$u({\bf r})$ the electric potential. 

With the aid of the above equations, the mean-field free energy, $ 
\Omega_{mf}$, which is the free 
energy function of Eq. \ref{free1} evaluated at the mean-field values of 
the densities and fields, can be put in the form
\begin{eqnarray}
-\Omega_{mf}&=& {kT\over v_h}\left({\cal Q}_l[w_h,w_t,u_h] 
+z_c{\cal Q}_c[u_c]
+z_s{\cal Q}_s[w_s]\right)+ E[\phi_h,\phi_t,\phi_s,\rho_h,\rho_c], \\
\label{mffreeenergy}
&=&kT(n_l+n_c+n_s)+E[\phi_h,\phi_t,\phi_s,\rho_h,\rho_c],
\end{eqnarray}
 with $E$ given by 
Eq. \ref{energy}.
The thermodynamic
potential, $\Omega$, is that appropriate to an incompressible system 
calculated in the
grand ensemble; the
negative of the osmotic pressure multiplied by the volume.
Thus the above equation states that 
the osmotic pressure is the sum of the ideal, partial 
osmotic pressures plus a correction due to the interactions.
Within mean field theory, this correction is simply the energy per unit 
volume of the system.

We now specify that the charges in the system can associate with or
disassociate from the headgroup in response to the local electrostatic
potential. This implies that the partition function of a single lipid,
${\cal Q}_l$ is to be averaged over the nominal charge distribution that
$Q_h=1$ with probability $p$, and $Q_h=0$ with
probability $1-p$ (Borukhov et al., 1998). The consequence of this averaging is
that ${\cal Q}_l[w_h,w_t,u_h]$ of Eq. \ref{lipidpf1} becomes
\begin{equation}
{\cal Q}_l[w_{h,eff},w_t]=\int \tilde{\cal
    D}{\bf r}_l
\exp\left\{-w_{h,eff}({\bf r}_l(1/2))-
\int_0^1ds w_t({\bf  r}_l(s))\right\},
\end{equation}
where
\begin{eqnarray}
w_{h,eff}({\bf r})&\equiv& w_h({\bf r})-\ln \int\tilde{\cal D}Q_h
\exp[Q_hu_h({\bf r})]\\          
                        &=&w_h({\bf r})-\ln [1+p(\exp[u_h({\bf r})]-1)].
\end{eqnarray}
Although this appears to introduce an unknown parameter $p$ into the
problem, the condition of charge neutrality,
\begin{equation}
\label{neutrality}
\int d{\bf r}[\rho_h({\bf r})+\rho_c({\bf r})]=0,
\end{equation}
 relates this parameter to the fugacity of the counter ions, $z_c$. In
 practice, we use this fugacity to control the pH and the amount of
 charge on the lipids.

There remains only to specify how the single-lipid partition function
is obtained. One defines the end-segment distribution function
\begin{equation}
q({\bf r},s)=\int{\cal D}{\bf r}_l(s)\delta({\bf r}-{\bf r}_l(s))
\exp\left\{-\int_0^s dt\left(\left[{1\over 
8R_g^2}|{d{\bf R}(t)\over dt}|^2\right]
+w_{h,eff}({\bf r}_l(t))\delta(t-1/2)+w_t({\bf r}_l(t))\right)\right\},
\end{equation}
which satisfies the equation
\begin{equation}
\label{endpoint} 
{\partial q({\bf r},s) \over\partial 
s}=2R_g^2\nabla^2q({\bf r},s)-
[w_{h,eff}({\bf r})\delta(s-1/2)+w_t({\bf r})]q({\bf r},s), 
\end{equation}
with initial condition  
\begin{equation}
q({\bf r},0)=1.
\end{equation}
The partition function of the lipid is then
\begin{equation}
\label{pfl}
{\cal Q}_l=\int d{\bf r}\ q({\bf r},1).
\end{equation}

From this expression for the single-lipid partition function and 
Eqs. \ref{phih}, \ref{phit}, and \ref{headcharge},  one
obtains expressions for the local density of the lipid heads
\begin{equation}
\label{lipidheads}
\phi_h({\bf r})=\exp[-w_{h,eff}({\bf r})]
q({\bf r},{1\over 2}-)q({\bf r},{1\over 2}-),
\end{equation}
of the lipid tails  
\begin{equation}
\label{lipidtails}
\phi_t({\bf r})=\int_0^1ds\ 
q({\bf r},s)q({\bf r},1-s)
\end{equation}
and of the charge density on the lipid heads
\begin{equation}
\label{lipidheadcharge}
\rho_h({\bf r})=-{p \exp[u_h({\bf r})]\over 1+p(\exp[u_h({\bf r})]-1)}
\phi_h({\bf r}).
\end{equation}

To summarize: there are four self-consistent equations to be solved
for the  fields $w_h$, $w_t$, $w_s$, and electrostatic potential
$u$. 
These equations, obtained from simple algebraic manipulation of 
Eqs. \ref{sc1} to
\ref{incompressibility},
 can be taken to be
\begin{eqnarray}
\gamma_tw_h({\bf r})-w_t({\bf r})&=&2\chi N[\gamma_t\phi_t({\bf r})
-\phi_h({\bf r})-\phi_s({\bf r})],\\
\gamma_sw_h({\bf r})-w_s({\bf r})&=&2\chi N(\gamma_s-1)\phi_t({\bf r})
+\lambda(\rho_c({\bf r})-\rho_h({\bf r})),\\
1&=&\phi_h({\bf r})+\gamma_t\phi_t({\bf r})+\gamma_s\phi_s({\bf r}),\\
R_g^2\nabla^2u({\bf r})&=-&\beta^*(\rho_h({\bf r})+\rho_c({\bf r})).
\end{eqnarray}
Note that we have chosen here to write the Poisson-Boltzmann equation,
Eq. \ref{pb}, in its local, rather than its integral form. When the
four fields are known, the corresponding densities follow from Eqs. 
\ref{phis}, \ref{rhoc2}, \ref{lipidheads}, \ref{lipidtails}, and
\ref{lipidheadcharge}.

Rather than attempt
to solve these equations in real space, a difficult task for the
periodic phases in which we are interested such as $H_{II}$, we recast
the equations in a form which makes straightforward their solution for a
phase of arbitrary space-group symmetry (Matsen and Schick, 1994).
\section{Theory: Fourier Space}
We note that the fields, densities, and the end point distribution
function depend only on one coordinate ${\bf r}$. Therefore in an
ordered phase, these functions reflect the space-group
symmetry of that phase. To make this symmetry manifest in the solution,
we expand all functions of position in a complete, orthonormal, 
set of functions,
$f_i({\bf r}),\  i=1,2,3...$, 
each of which have the desired space group symmetry; {\em e.g.} 
\begin{eqnarray}
\phi_h({\bf r})&=&\sum_i\phi_{h,i}f_i({\bf r}),\\
\delta_{i,j}&=&{1\over V}\int d{\bf r}f_i({\bf r})f_j({\bf r}).
\end{eqnarray}
Furthermore we choose the $f_i({\bf r})$ to be eigenfunctions of the
Laplacian
\begin{equation}
\nabla^2f_i({\bf r})=-{\lambda_i\over D^2}f_i({\bf r}),
\end{equation} where $D$ is a length scale for the phase.
The functions for the lamellar phase are clear. They can be taken to be
\begin{eqnarray}
f_1({\bf r})&=&1,\\
f_i({\bf r})&=&\sqrt{2}\cos[2\pi (i-1) x/D],\qquad i\ge 2,
\end{eqnarray}
 Expressions for the unnormalized basis functions for other space-group 
symmetries can be found in X-ray tables
(Henry and Lonsdale, 1969) as they are intimately related to the Bragg peaks.
In the tables cited, those for the hexagonal phase, space group
(p6m) can be found on page 372, and that of the bcc phase, space group (Im3m) 
 on page 524.

The four self-consistent equations become
\begin{eqnarray}
\label{sceq1}
\gamma_tw_{h;i}-w_{t;i}&=&2\chi N[\gamma_t\phi_{t;i}-\phi_{h;i}-\phi_{s;i}],\\
\gamma_sw_{h;i}-w_{s;i}&=&2\chi
N(\gamma_s-1)\phi_{t;i}+\lambda(\rho_{c;i}-\rho_{h;i}),\\
\delta_{1,i}&=&\phi_{h;i}+\gamma_t\phi_{t;i}+\gamma_s\phi_{s;i},\\
\label{sceq4}
{\lambda_iR_g^2\over D^2}u_i&=&\beta^*(\rho_{h;i}+\rho_{c;i})
\end{eqnarray}
To obtain the partition functions and 
densities, we proceed as follows. For any function 
$G({\bf r})$, we can define a symmetric matrix 
\begin{equation}
(G)_{ij}\equiv{1\over V}\int f_i({\bf r})G({\bf r})f_j({\bf r})d{\bf r}
\end{equation}
Note that $(G)_{1i}=(G)_{i1}=G_i$, the coefficient of $f_i({\bf r})$ in
the expansion of $G({\bf r})$. Matrices corresponding to functions of
$G({\bf r})$, such as 
\begin{equation}
\left(e^G\right)_{ij}\equiv{1\over V}
\int f_i({\bf r})e^{G({\bf r})}f_j({\bf r})d({\bf r}),
\end{equation}
are evaluated by making an orthogonal transformation which diagonalizes 
$(G)_{ij}$.
With this definition, Eqs. 
\ref{phis} and \ref{rhoc2} yield the solvent density and counter ion
charge density
\begin{eqnarray}
\phi_{s;i}&=&z_s\left( e^{-w_s}\right)_{i,1},\\
\rho_{c;i}&=&z_c\left( e^{-u_c}\right)_{i,1},\nonumber\\
          &=&z_c\left( e^{-(u-u_s)}\right)_{i,1}.\\
\end{eqnarray}
To obtain the remaining densities, we
need the end-point distribution
function. From Eq. \ref{endpoint} we obtain
\begin{eqnarray}
{dq_i(s)\over ds}&=&
-\sum_j[A_{ij}+(w_{h,eff})_{ij}\delta(s-1/2)]q_j(s),\\
A_{ij}&=&{2R_g^2\over D^2}\lambda_i\delta_{ij}+
(w_{t})_{ij},
\end{eqnarray}
with initial condition $q_i(0)=\delta_{i,1}$. The solution of this 
equation is
\begin{eqnarray}
q_i(s)&=&\left(e^{-A s}\right)_{i,1},\qquad {\rm if}\ s<1/2\nonumber \\
      &=&\sum_j\left(e^{-w_{h,eff}}\right)_{ij}\left(e^{-A/2}\right)_{j,1}
,\qquad s=1/2
\nonumber \\
      &=&\sum_{j,k}\left(e^{-A(s-1/2)}\right)_{i,j}
\left(e^{-w_{h,eff}}\right)_{jk}\left(e^{-A/2}\right)_{k,1}
, \qquad s>1/2.
\end{eqnarray}
From this, the remaining  densities follow from Eqs. \ref{lipidheads}, 
\ref{lipidtails}, and \ref{lipidheadcharge}:
\begin{eqnarray}
\phi_{h;i}&=&\sum_{jkl}\left(e^{-w_{h,eff}}\right)_{ij}
\Gamma_{jkl}q_k({1\over2}-)q_l({1\over2}-),\\
\phi_{t;i}&=&\int_0^1 ds\sum_{jk}\Gamma_{ijk}q_j(s)q_k(1-s),\\
\rho_{h;i}&=&-\sum_{j}\left({pe^{u_h}\over 
1+p(e^{u_h}-1)}\right)_{i,j}\phi_{h;j},
\end{eqnarray}
with
\begin{equation}
\Gamma_{ijk}\equiv{1\over V}\int f_i({\bf r})f_j({\bf r})f_k({\bf r}).
\end{equation}
 The mean-field free energy, Eq. \ref{mffreeenergy},
takes the form 
\begin{equation}
-\Omega_{mf}={kTV\over v_h}(\phi_{t;1}+\phi_{s;1}+\rho_{c;1})+E,
\end{equation}
with the mean-field energy being given by
\begin{equation}
E={kTV\over v_h}\sum_i[2\chi N(\phi_{h;i}+\phi_{s;i})\phi_{t;i}+
{1\over
  2}(\rho_{h;i}+\rho_{c;i})u_i-\lambda(\rho_{c;i}-\rho_{h;i})\phi_{s;i}].
\end{equation}
 We have expressed the Coulomb energy as a product of the
 charge densities and electrostatic potential. Note that this free
energy still depends parametrically on $D$, the length scale of the phase, 
so that the value of $D$ which minimizes it must be determined.
Once this is done, we compare  the free
 energies obtained for phases of different space-group symmetry, and thereby
 determine the phase diagram of our model lipid system.

The infinite set of self-consistent equations, Eqs. \ref{sceq1} to \ref{sceq4}
must be truncated in order to be solved numerically. 
We have employed up to 
50 basis functions. This truncation is 
sufficient to ensure, for $T^*>0.03$ and $1/(1+\gamma_t)<0.66$, 
an accuracy of $10^{-4}$ in 
the free energy $v_h\Omega_{mf}/kTV$.
As noted, one must also determine the length scale which minimizes 
the free energy. This is usually straightforward as there is a single 
well-defined minimum for a phase of given symmetry at given thermodynamic 
parameters; temperature, and chemical potentials. Were there more than one 
minimum, this would reflect a tendency for the system to phase separate into
two phases with the same non-trivial space-group symmetry, an extremely 
unusual occurrence. The single minimum that one finds normally is sharp,
that is, the free energy varies rather rapidly with $D$. Only in cases in
which phases are greatly swollen is the minimum extremely shallow and 
difficult to locate.

\section{Results: The Neutral Lipid}
  We first apply our method to a neutral lipid. 
 We
  show here the phase behavior of the  neutral
  lipid, in the absence and in the presence of solvent. Figure 1 shows
  the phase diagram of the pure lipid as a function of the
 dimensionless  temperature $T^*$,
  and the architecture of the lipid. The latter is
  characterized by the single parameter $1/(1+\gamma_t)$ which is the 
 relative volume of the headgroup to that of the
  entire lipid. It is analogous to, but not the same as, the single
  parameter used by Israelachvili to characterize the geometry of lipids
  (Israelachvili, 1985). Shown are the lamellar phase,
  $L_{\alpha}$, the normal and inverted hexagonal phases, $H_{I}$ and
  $H_{II}$, and the normal and inverted body-centered cubic phases
  $bcc_{I}$ and $bcc_{II}$. The occurrence of bicontinuous phases will be 
  discussed in a later paper.

  One sees that the phase behavior
  is reasonable and in accord with packing considerations;
   as the headgroup increases in volume, the
  system passes through a series of phases from the inverted ones with
  most curvature, through the lamellar phase, to the normal ones
  of most curvature. We also note that as the temperature increases,
   the
  lamellar phase becomes unstable to one or the other of the 
  hexagonal phases. This is very
  clear for the $H_{I}$ phase, less so for the $H_{II}$ phase, but is
  true for the latter over much of the temperature range 
  as the slope of the $L_{\alpha}/H_{II}$ phase
  boundary is slightly positive below $T^*$  of about 0.045. This
  transition with temperature 
  is well known in the analogous case of diblock copolymers
  (Leibler, 1980). In order to model a lipid which, like
  phosphatidylserine, adopts the $H_{II}$ configuration when essentially
  neutral (Cullis et al. 1985, Bezrukov et al., 1999), 
  we have chosen $1/(1+\gamma_t)=0.24$ in
  our subsequent studies. For comparison, the value appropriate for 
  DOPE, calculated from the molecular
  volumes in the literature (Rand and Fuller, 1994) is $0.254$. For the 
convenience of the
reader interested in carrying out similar calculations, we present in Table I 
the values of the first three non-trivial Fourier components of the headgroup
density $\phi_h({\bf r})$, the lattice parameter $D/R_g$, and the free energy 
$\Omega_{mf}v_h/kTV$ for the $L_{\alpha}$, $H_{II}$ and $bcc_{II}$ phases at
 $T^*=0.04$. We note, in passing, that the relative intensities of
 X-ray Bragg peaks can be determined directly from the Fourier components
of the various densities with which they are associated.  

  The effect on this neutral lipid of the addition of a solvent of small
  volume, 
  characterized by $\gamma_s=v_s/v_h=0.1$, close to the value of
  $0.096$ (Rand and Fuller, 1994;  Kozlov et al., 1994)
 appropriate to water and a phosphatidylethanolamine
  headgroup,  is shown in Fig. 2. 
  There is a lamellar phase at small solvent volume fractions and low
  temperatures. This phase becomes unstable with 
  respect 
  to the  $H_{II}$ phase which envelops it at higher temperatures.
  There is a large region of two-phase coexistence between the
  ordered lipid-rich phases and an  almost pure solvent phase. These
  features are reasonable, and are observed in the systems of aqueous dialkyl
  didodecylphosphatidylethanolamine and of 
   diacyl diarachinoylphosphatidylethanolamine  (Seddon et al, 1984). Of
   particular interest is that we find a small temperature region 
   of re-entrant
   hexagonal-lamellar-hexagonal transitions, an unusual feature which
   has been observed in DOPE 
   (Gawrisch et al. 1992, Kozlov et al. 1994). As a consequence, there is an 
azeotrope at which the transition between lamellar and hexagonal phases 
occurs without a change in the concentration of water. We have used the
 coordinates
 of this point, $T^*=0.06$ and $\phi_s=1.42$, 
denoted $T_0$ and $(\phi_s)_0$, 
 to 
   normalize the temperature and solvent-density axes.   
   There is  a small region of $bcc_{II}$ in our phase
   diagram. Again, the possible occurrence of bicontinuous phases will be 
   examined in a later publication. 
  The uncertainty in the temperature of the phase boundaries, $\delta T/T$ 
  introduced by the truncation of the number of basis functions, 
  is approximately $2\times 10^{-3}$.
  
As the volume fraction of water is increased, we find that the period of all 
structures increases, 
as is expected. In Fig. 3, we compare experimental 
results  on DOPE taken in the inverted hexagonal phase at a temperature 
$T=22^{\circ} C$, 
just above that of the azeotrope (Rand and Fuller, 1994), to our  
values calculated just above the azeotrope. Knowing the molecular weight of
 DOPE, we convert the volume fraction of solvent, $\gamma_s\phi_s$, which
 occurs in the
 calculation, to the experimental variable of weight fraction of water. 
The lattice parameter of the hexagonal phase in the calculation, however, 
is measured in units
of the radius of gyration of either lipid tail. What value should be taken to 
model DOPE is unknown. 
Hence we have used in the comparison the lattice parameter $D$, in units of 
$D_0$, the lattice parameter at the azeotrope.
 The agreement is excellent.

An effective value of the radius of gyration can be defined as that value
 which brings agreement between the calculated and measured lattice parameters.
As the former, at the azeotrope, is  $D(T_0)\equiv D_0=4.79 R_g^0$, 
and the latter is
58.9\AA\ (Rand and Fuller, 1994), the equivalent radius of gyration at the 
temperature of the azeotrope, $R_g^0$, is 
12.3\AA\ for a single tail,
not unreasonable when compared to the extended length of a single chain 
of DOPE which is  approximately 26\AA.        

 As the temperature of the system is lowered, the period of all
  structures increases, which is due to the lengthening of the tails as
  their entropy decreases. We would again like to  compare our results with 
those of the DOPE system. In order to do so, we must address the temperature
 dependence of the radius of gyration, $R_g$, which appears in the theory,
and which is not given {\it a priori}. Because the model chain is flexible,
the radius of gyration is related to the mean square end-to-end distance,
 $\bar R$, by a numerical constant, $R_g=\bar R/\sqrt{6}$, so their 
 dependence on temperature is the same. 
To compare our results to DOPE, we shall assume that the temperature
 dependence of the radius of gyration which appears in the calculation is
 the same as that given for lipid chains by 
the Rotational Isomeric States Model, 
a model 
 which describes the properties of such chains very well (Flory, 1969;
 Mattice and Suter, 1994). Thus we assume
\begin{equation}
R_g(T)=c\left({1-<\cos \phi>\over 1+<\cos\phi>}\right)^{1/2},
\end{equation}
where the angle $\phi$ takes the values $180^\circ$ and $\pm70^{\circ}$ 
corresponding to $trans$, $gauche^+$ and $gauche^-$ configurations, and
 $c$ is a constant. The 
statistical average of $\cos\phi$ is
\begin{equation}
<\cos\phi>={\cos (180^{\circ})+\sigma\cos (70^{\circ})
+\sigma\cos(-70^{\circ})  
            \over 1+2\sigma}
\end{equation}
with $\sigma=\exp(-T_{rism}/T)$ and $T_{rism}=280.25^{\circ}$K.
From the behavior with temperature of $R_g(T)$, 
the lattice parameter, $D(T)$, at any temperature can be obtained 
from
\begin{equation}
{D(T)\over D(T_0)}={[D(T)/R_g(T)]\over [D(T_0)/R_g(T_0)]}
{R_g(T)\over R_g(T_0)}.  
\end{equation}
Again, it is the factor $D(T)/R_g(T)$ which occurs naturally in the 
calculation.

A comparison of the experimentally measured (Tate and Gruner, 1989) 
and theoretically calculated 
lattice parameters  {\it vs.} temperature is shown in 
Fig. 4. In part (a), the variation of the parameter of the $H_{II}$ 
is shown at two different lipid weight
fractions. The agreement is very good. In part (b), the comparison is made 
of the $H_{II}$ and $L_{\alpha}$ parameters along the coexistence with 
excess water. Note that this comparison is a much more stringent test, for it  
requires not  only that the dependence of the lattice parameters on solvent
concentration and on temperature be reproduced well by the calculation, but
also that the phase boundaries be given well. Considering these requirements, 
the agreement is rather good. It should be noted that the agreement in 4(b)does
 not depend on the exact temperature of the triple point, which is 
difficult to locate precisely, but only on the existence of stable $H_{II}$ and
$L_{\alpha}$ phases which coexist with excess water.

As seen in Fig. 4(b), the lattice parameter of the $H_{II}$ phase is much
 larger than that of the $L_{\alpha}$ at the triple point.
This
 is due to the coexistence with 
 excess solvent, which swells the hexagonal cores, but which is only weakly 
 present between the lamellae. In contrast, when the lamellar and 
 hexagonal phases are only in two-phase coexistence with one another and 
 there is no excess solvent, the hexagonal 
 phase in general has a smaller lattice parameter than the coexisting 
 lamellar phase, as shown in Fig. 5. This is due to the fact that over 
 almost all of their coexistence region, the $H_{II}$ phase has a smaller 
volume fraction of solvent than does the $L_{\alpha}$ phase, as can be 
seen from the phase diagram of Fig. 2. Only over the re-entrant 
region, which occurs as the triple point is approached, does this balance 
shift. This shift in the relative size of the latter parameters through
the re-entrant region is in agreement with experiment (Kozlov et al., 1994). 

  It is of interest to determine if any one effect can be said to drive
the transition from the $H_{II}$ to $L_{\alpha}$ phase in the neutral
system. To this end we examine the individual terms in the 
thermodynamic potentials  per unit volume 
$\Omega_{mf} v_h/kTV=Ev_h/kTV-S_lv_h/kV-(S_sv_h/kV+\phi_s\ln z_s)$ of the
$L_{\alpha}$ and $H_{II}$ phases as the transition is crossed by
increasing the solvent fugacity, $z_s$, at constant temperature 
$T/T_0=0.67$. In Table II we show the contributions to the free energy
per unit volume of the $L_{\alpha}$ phase and that of the $H_{II}$ phase
coming from the interaction energy, $Ev_h/kTV$, the lipid tails,
$-S_lv_h/kV$, and the solvent $-S_sv_h/kV-\phi_s\ln z_s.$ All
contributions are evaluated at the transition itself, which occurs at
$z_s\approx 3.14$, and are measured with respect to the free energy per
unit volume of the disordered phase.  We also show the difference in the
contribution of each term to the free energies of each phase, and the
derivative of this difference with respect to the solvent fugacity.  The
difference in the contribution of the entropy of the lipid tails is
positive because, with the lipid architecture we have chosen, the large
tail volume relative to that of the head favors the hexagonal phase. The
interaction energy favors the lamellar phase, as does the solvent,
presumably because the interstices of that phase are two-dimensional,
while those of the inverted hexagonal phase are one-dimensional.  The
difference between the lipid entropy contributions decreases with
increasing solvent concentration because the packing constraints in the
$H_{II}$ phase become more severe as the size of the cores increases
(Gruner, 1989). However it is apparent that neither this term, nor
either of the others, changes so rapidly with solvent concentration 
compared to the others that
any particular effect can be said to ``drive'' this transition.

\section{Results: The Charged Lipid}
\subsection{Coulomb Interactions Only}
We now turn on the negative charge of the headgroups by varying the
chemical potential of the free counter ions while enforcing charge
neutrality. Increasing the density of free, positive, counter ions in our
closed system is equivalent to increasing the magnitude of the negative
charge density on the headgroups. It therefore corresponds to an
increase in the pH of an experimental system.  The charge on the 
headgroups is annealed, meaning that it is determined by the local value of
the electrostatic potential, and therefore the headgroup charge varies
with the location of that group. The parameter $\beta^*$, defined in
Eq. \ref{betastar}, measures the strength of the Coulomb
interaction. It can be written as the ratio of two lengths, 
$\beta^*=\xi/L_l$, where $\xi\equiv e^2/\epsilon kT$ is the Bjerrum 
length, and $L_l\equiv v_h/4\pi R_G^2$ is a length characterizing the 
architecture of the lipid. It is reasonable that $\beta^*$ be larger than 
or of order unity, and we have arbitrarily taken $\beta^*=1.$  

We find for this system that the effect of increasing
headgroup charge is to decrease the temperature of all transitions.  As
an example, we show in Fig. 6 the way the temperature, $T^*$,
of the transition between $L_{\alpha}$ and $H_{II}$ phases varies with
the magnitude of the average charge density of the headgroups 
${\bar\rho}_h\equiv\rho_{h;1}$.  The
range of charge density corresponds to the headgroups varying from being
neutral to fully charged. The region beyond the almost vertical line at
$|{\bar\rho}_h|\approx 0.24$ would correspond to the headgroups being charged
with a probability greater than unity, and is therefore unphysical. The
maximum value of $|\bar{\rho}_h|$ is slightly different in the two
phases. Over the physical range, one sees a 10 \% decrease in the
transition temperature, and therefore, a stabilization of the $H_{II}$
phase with respect to the $L_{\alpha}$ phase.  Although at first
surprising, this is a reasonable result, and can be understood as
follows. In the neutral system, the single interaction parameter in the
system, $\chi$, is a measure of the tendency for the system to order,
and is proportional to the inverse temperature.  When the charges are
turned on, the Coulomb repulsion between headgroups opposes ordering,
and therefore has an effect similar to a decrease in $\chi$ or an
increase in temperature.
 As we have noted earlier, increasing the temperature
causes the lamellar phase to become unstable to the hexagonal phase.
A complimentary view is to note that increasing the Coulomb repulsion
between headgroups increases the area per headgroup, and thus 
the area between headgroups, 
an effect which,
again, is equivalent to increasing the temperature 
(Seddon and Templer. 1995). The increase in area
 permits the tails to splay further out,
favoring the formation of the inverted hexagonal phase.
The very same result of increasing the density of charges in a system is
seen in the lyotropic phases of polyelectrolytes. There one
finds that the phase diagram depends only on the ratio of $\chi$ to the
charge on the polymer, at least in the absence of fluctuations (Nyrkova
et al., 1994). Hence an increase in the charge is equivalent to a
decrease in $\chi$ or an increase in temperature, leading to an
instability of the lamellar phase.

We are confident, therefore, that the transition from $L_{\alpha}$ to
$H_{II}$ phases with increasing charge on the headgroups is the
correct result of the model. However it is completely opposite to the 
experimental results on phosphatidylserine in water (Hope and Cullis,
1980, Bezrukov et al., 1999). Hence there must be a crucial physical 
mechanism which is not
included in the model which includes  Coulomb interactions only 
between objects with a net charge. One such mechanism 
is the short-ranged attraction between the charges of the headgroup and 
the dipoles of the neutral water molecules which will cause the latter to
aggregate around the former.
Furthermore the counterions attracted to the head group  will now also
be associated with waters of hydration endowing them with a
 non-negligible volume.  
The additional volume of all these waters will
 increase the effective volume of the
headgroup, and therefore tend to stabilize the lamellar phase.
 This additional volume will increase with the charge of the
headgroup, and could be sufficient to counteract the
Coulomb repulsion tending to stabilize the hexagonal phase.
This competition is investigated next.

\subsection{ Coulomb and Short-Ranged Solvent
  Interactions}
We now include in the system an attractive contact interaction between
charges and the neutral solvent. This models the short-ranged, thermally
averaged, interaction between charges and the water dipole, one which
varies with separation $r$ as
$(kT/6)(u/e\xi)^2(\xi/r)^4\equiv\omega(r)$, with $u$ the dipole moment
of water and $\xi$ the Bjerrum length. The
above expression is valid for distances such that
$r/\xi>(u/e\xi)^{1/2}.$ For water, $\xi\approx 7$ \AA, and
$\omega(r)/kT\approx 4.6\times 10^{-4}(\xi/r)^4$. To approximate this
short-ranged interaction by a contact interaction of dimensionless
strength $\lambda$ is equivalent to employing $\omega/kT$ evaluated at
some fixed distance. Any reasonable choice shows that $\lambda$ is
small. We have arbitrarily chosen $\lambda=0.1$, which corresponds to
$\omega(r)/kT$ evaluated at $1.9$\AA, a distance within the regime in
which the approximate expression for the charge-dipole interaction is
valid.

The results for the system with the Coulomb interaction of strength
$\beta^*=1$ and which includes the interaction between charges and
solvent of strength $\lambda=0.1$ are qualitatively different from
those of the charged lipids without this interaction.  The temperatures
of all transitions between ordered phases now increase with increasing
headgroup charge, whereas previously they decreased. Shown in Fig. 7 is
the temperature of the transition between $H_{II}$ and $L_{\alpha}$
phases as a function of the magnitude of the average charge density on
the headgroups.  One sees that with increasing headgroup charge, the
inverted hexagonal phase becomes unstable with respect to the lamellar
phase, just as in the experiments on phosphatidylserine in water (Hope
and Cullis, 1980, Bezrukov et al., 1999)

In Fig. 8a we show the volume fraction profiles in the $L_{\alpha}$ phase
at a value of
$z_c=0.1228$, $z_s=3$, and temperature $T^*=0.04$ at which the 
$H_{II}-L_{\alpha}$ transition occurs. 
The position through the system is divided by the
lattice parameter of the phase which, in units of the radius of gyration of the
lipid, is $D_{L}/R_g\approx\ 4.14.$ All looks reasonable. In particular, 
the volume
fraction of solvent within the bilayer is negligible. This is true also
when the lipid is neutral. In Fig. 8b, the
charge density profile of the same structure is shown. One sees that the
charge on
the headgroup mimics, but does not reproduce, the headgroup volume
fraction. This is because the charge on the headgroup is not fixed, but
varies with the local electrostatic potential. 
The counter ion density is fairly uniform because we have
employed a single dielectric constant, that of water, throughout the
system. Were we to require that the dielectric constant be
position-dependent, and to take the much lower value in the tail region
appropriate to it, the counter ion density there would be much reduced. 
In Figs. 9a and 9b we show
the same quantities for the $H_{II}$ phase at the same value of
$z_c$. The cut through the system is taken along
the nearest-neighbor direction, 
and the distance is normalized to its lattice parameter
 $D_{H}\approx 3.83R_g.$
Again the wavelength of the $H_{II}$ phase is smaller than that of the 
$L_{\alpha}$ phase because, in two-phase coexistence, {\em i.e.} in the
absence of a reservoir of excess water, the cores of the cylinders are
not swollen with water and the hexagonal phase contains a smaller volume
fraction of water than does the lamellar phase. One can infer from
Fig. 9a that in the nearest neighbor direction, there is far more
interdigitation of lipid tails than in the lamellar phase. This makes
sense, as the tails must certainly stretch to fill the space between
cores in the second-neighbor direction, so that interdigitation is 
expected in the nearest-neighbor direction.
 
To investigate this transition further, we show in Table III the
contributions in the $L_{\alpha}$ phase and in the $H_{II}$ phase of the
various terms in the thermodynamic potential per unit volume
$\Omega_{mf} v_h/kTV=(E_1+E_2+E_3)v_h/kTV-S_lv_h/kV-(S_sv_h/kV+\phi_s\ln z_s)-
(S_cv_h/kV+\rho_c\ln z_c)$. Here $E_1$ is the hydrophilic-hydrophobic
 interaction
proportional to $\chi N$, $E_2$ is the electrostatic interaction
proportional to $\beta^*$, and $E_3$ is the charge-solvent interaction
proportional to $\lambda$. These contributions are evaluated at the
transition itself, and are measured from the free energy per unit volume
of the disordered phase. We also show the difference between these
contributions to each phase, and the derivatives of each of these
differences with respect to the counter ion fugacity, $z_c$. There are
several interesting things to note. The electrostatic energy is a
relatively small contribution to the free energy of each phase, and
hardly differs between them.  Therefore it does not have a large effect
in bringing about the transition.  The contribution of the counter ions
to the free energy of each phase is of the same order of magnitude as the
electrostatic interaction and, like it, does not change rapidly with the
counter ion fugacity. The contribution of the short-range charge-solvent
interaction is small, but it changes most rapidly with the counter ion
fugacity, and therefore appears to be most important in actually
bringing about the transition itself.

The physical mechanism in the experimental systems appears now to be
clear. The lipid with an almost neutral headgroup forms the $H_{II}$
phase because the volume of the headgroup is relatively small compared
to that of the entire lipid. As the charge on the headgroup is turned
on, it attracts an increasing volume of waters of hydration via the
attractive interaction between the charge and the dipoles of water. 
In addition, more counter ions, enlarged by their own waters of hydration,
are attracted to the headgroup.
Thus the
headgroup becomes effectively larger, and drives the transition to the
$L_{\alpha}$ phase. In doing so, it overcomes the effect of the Coulomb
repulsion between headgroups which, in fact, opposes the transition to
the $L_{\alpha}$ phase.

We have argued earlier that an increase in  Coulomb repulsion was
similar to an increase in temperature,  and
therefore its effect in favoring the $H_{II}$ phase could be understood
from the phase diagram of Fig. 1. In a similar way, an increase in the
strength of the charge-solvent interaction increases the volume
fraction of the headgroup, so that one moves to the right in Fig. 1 and
stabilizes the lamellar phase. The combined effect of increasing the
headgroup charge is to increase the
effective volume fraction of the 
headgroup and effective temperature. 
The phase which is ultimately stabilized results from the
competition of these two effects. By changing the strength of the
Coulomb interaction, we have verified that one can alter the effect of
this competition, and not only stabilize the inverted hexagonal phase,
but also bring about the inverted b.c.c. phase or even the disordered
phase, scenarios which are again understandable from our discussion and
the phase diagram of Fig. 1. But that the result of this  competition 
can be the stabilization of the lamellar phase at the expense of the
inverted hexagonal phase has been demonstrated by our model calculation,
as well as by experiment.

We gratefully acknowledge useful communications with Drs. Pieter Cullis,
John Seddon, and Sol Gruner. 
This work was supported in part by the National Science Foundation under
grant numbers DMR9531161 and DMR9876864. One of us, (MS), would like to
 thank the CEA, Saclay, for their gracious hospitality while 
 this paper was being written.
\newpage
\begin{center}
{\bf REFERENCES}
\end{center}
\begin{description}
\item{} Ben-Shaul, A., I. Szleifer, and W.M. Gelbart. 1985. Chain organization
and thermodynamics in micelles and bilayers: I theory. {\em
  J. Chem. Phys.} 83:3597-3611.
\item{} Bezrukov, S.M., R.P. Rand, I. Vodyanoy, and V.A. Parsegian. 1999.
Lipid packing stress and polypeptide aggregation: alamethicin channel
probed by proton titration of lipid charge. Faraday Discuss. 111:000-000
\item{} Borukhov, I., D. Andelman, and H. Orland. 1998. Random
  polyelectrolytes and polyampholytes in solution. {\em Eur. Phys. J. B} 
5:869-880.
\item{} Cullis, P.R., M.J. Hope, B. de Kruijff, A.J. Verkleij,
  and C.P.S. Tilcock. 1985. Structural properties and functional roles
  of phospholipids in biological membranes. {\em in} 
  Phospholipid and Cellular Regulation, vol. 1., J.F. Kuo (ed.). CRC
  Press, Boca Raton, Florida 1-59.
\item{} Fattal, D.R. and A. Ben-Shaul. 1994. Mean-field calculations of chain
packing and conformational statistics in lipid bilayers: comparison with
experiments and molecular dynamics studies. {\em Biophys. J.}
67:983-995.
\item{} Flory, P.J. 1969 Statistical mechanics of chain molecules, Wiley
  Interscience, New York.
\item{}Gawrisch, K., V.A Parsegian, D.A. Hadjuk, M.W. Tate, S.M. Gruner,
  N.L. Fuller, and R.P. Rand. 1992. Energetics of a
  hexagonal-lamellar-hexagonal-phase transition sequence in 
dioleoylphosphatidylethanolamine membranes. {\em Biochemistry}
31:2856-2864.
\item{} Gruen, D.W.R. 1981. A statistical mechanical model of the lipid
  bilayer above its phase transition {\em Biochim. Biophys. Acta}
  595:161-183
\item{} Gruen, D.W.R. 1985. A model for the chains in amphiphillic
  aggregates.1. Comparison with a molecular dynamic simulation of a
  bilayer {\em J. Phys. Chem.} 89:146-153.
\item{} Gruner, S.M. 1989. Stability of lyotropic phases with
  curved surfaces. {\em J. Phys. Chem.}, 93:7652-7570.
\item{} Helfrich, W. 1973. Elastic properties of lipid bilayers: theory
  and possible experiments. {\em Z. Naturforsch}. 28C:693-703.
\item{}Heller, H., M. Schaefer, and K. Schulten. 1993. Molecular dynamics 
simulation of a bilayer of 200 lipids in the gel and in the liquid-crystal 
phases. {\em J. Phys. Chem.} 97:8343-8360.
\item Henry, N.F.M. and K. Lonsdale 1969 (eds.) International Tables
for X-Ray Crystallography, Kynoch, Birmingham. 
\item{} Hope, M.J., and P.R. Cullis. 1980. Effects of divalent
  cations and pH on phosphatidylserine model membranes: A $^{31}$P NMR
  study. {\em Biochemical and Biophysical Research Communications}
  92:846-852.
\item{} Israelachvili, J.N. 1985. Intermolecular and Surface Forces,
  Academic
  Press, San Diego.
\item{} Jacobs, R.E. and S.H. White. 1989. The nature of the hydrophobic
  bonding of small peptides at the bilayer interface: implications for
  the insertion of transbilayer helices. {\em Biochemistry} 28:3421-3437
\item{} Kirk, G.L., S.M. Gruner, and D.L. Stein. 1984 A thermodynamic
  model of the lamellar to inverse hexagonal phase transition of lipid
  membrane-water system. {\em Biochemistry}. 23:1093-1102.
\item{} Kirk, G.L. and S.M. Gruner. 1985. Lyotropic effects of alkanes
  and headgroup composition on the $L_{\alpha}-H_{II}$ lipid liquid
  crystal phase transition: hydrocarbon packing {\em versus} intrinsic
  curvature. {\em J. Physique} 46:761-769.
\item{} Kozlov, M.M., S. Leiken, and R.P. Rand. 1994.
  Bending, hydration
  and interstitial energies quantitatively account for the
  hexagonal-lamellar-hexagonal re-entrant phase transition in
  dioleoylphosphatidylethanolamine. {\em Biophys. J} 67:1603-1611.
\item{} de Kruijff, B. 1997. Lipid polymorphism and biomembrane
  function. {\em Current Opinion in Chemical Biology},1:564-569.
\item{} Leermakers, F.A.M., and J.M.H.M. Scheutjens. 1988 Statistical
  thermodynamics of association colloids. I. Lipid bilayer
  membranes. {\em J. Chem. Phys.} 89:3264-3274.
\item{} Leermakers, F.A.M., J.M.H.M. Scheutjens, and
  J. Lyklema. 1990. Statistical thermodynamics of association
  colloids. IV. Inhomogeneous membrane systems. {\em
    Biochim. Biophys. Acta} 1024:139-151.
\item{} Leibler, L. 1980. Theory of Microphase Separation in Block Copolymers. 
{\em Macromolecules} 13:1602-1617.
\item{} Marcelja, S. 1974. Chain ordering in liquid crystals
  II. Structure of bilayer membranes {\em Biochim. Biophys. Acta} 367:165-176
\item{} Matsen, M.W., and M. Schick. 1994. Stable and unstable
  phases of a diblock copolymer melt. {\em Phys. Rev. Lett.}
  72:2660-2663. 
\item{} Matsen, M.W. 1995. Stabilizing new morphologies by
  blending homopolymer with block-copolymer. {\em Phys. Rev. Lett.}
  74:4225-4228.
\item{} Mattice, W.L. and U.W. Suter. 1994 Conformational theory of
  large molecules; the rotational isomeric state model in macromolecular
  systems, Wiley-Interscience, New York.
\item{} Meijer, L.A., F.A.M. Leermakers, and A. Nelson. 1994. Modelling
  of electrolyte ions-phospholipid layers interaction {\em Langmuir} 
10:1199-1206.
\item{} M{\"u}ller, M. and M. Schick, 1998. Calculation of the
  phase behavior of lipids. {\em Phys. Rev. E} 57:6973-6978.
\item{} Nyrkova, I.A., A.R. Khokhlov,.and
  M. Doi. 1994. Microdomain structures in polyelectrolyte systems:
  calculation
of the phase diagram by direct minimization of the free energy. {\em
  Macromolecules} 27:4220-4230.
\item{} Rand, R.P. and N.L. Fuller. 1994. Structural dimensions and their 
changes in a reentrant hexagonal-lamellar transition of phospholipids.
{\em Biophys. J.} 66: 2127-2138.
\item{} Rand, R.P. and V.A. Parsegian. 1989. Hydration forces between
  phospholipid bilayers. {\em Biochim. Biophys. Acta.} 1031:1-69.
\item{} Seddon, J.M., G. Cevc, and D. Marsh. 1983. Calorimetric studies
  of the gel-fluid ($L_{\alpha}-L_{\beta}$) and lamellar-inverted
  hexagonal ($L_{\alpha}-H_{II}$) phase transitions in dialkyl- and
  diacylphosphatidylethanolamines. {\em Biochemistry}, 22:1280-1289.
\item{} Seddon, J.M., G. Cevc, R.D. Kaye, and
  D. Marsh. 1984. X-ray diffraction study of the polymorphism of
  hydrated diacyl- and dialkylphosphatidylethanolamines. {\em
    Biochemistry} 23:2634-2644.
\item{} Seddon, J.M. 1990. Structure of the inverted hexagonal
  ($H_{II}$) phase, and non-lamellar phase transitions of lipids. {\em
    Biochimica and Biophysica}, 1031:1-69.
\item{} Seddon, J.M. and R. H. Templer 1995. Polymorphism of lipid-water
  systems {\em in} Structure and dynamics of membranes, vol. 1,
  R. Lipowsky and E. Sackmann (eds.). Elsevier Press, Amsterdam 97-160.
\item{} Steenhuizen,L., D. Kramer, and
  A. Ben-Shaul. 1991. Statistical thermodynamics of molecular
  organization in the inverse hexagonal phase.{\em J. Phys. Chem.} 
95:7477-7483.
\item{} Tate, M.W. and S.M. Gruner 1989.  Temperature dependence of the 
structural dimensions of the inverted hexagonal ($H_{II}$) phase of
phosphatidylethanolamine-containing membranes. {\em Biochemistry} 
28:4245-4253.
\end{description}


\newpage
\begin{description}
\item[Table I] Anhydrous, neutral lipid: the lattice parameter, 
the free energy,
and the first 3 non-trivial Fourier components 
of $\phi_h({\bf r})$ for $L_\alpha$,
$H_{II}$ and $bcc_{II}$ phases at $T^*=0.04$ and $1/(1+\gamma_t) = 0.24$.
Recall that $\phi_{h,1} = \phi_{t,1} = 0.24$ for all phases.
\end{description}
\bigskip
$
\begin{array}{lccccc}
&D/R_g&v_h \Omega_{mf}/kTV&\phi_{h,2}&\phi_{h,3}\times 10^2
&\phi_{h,4}\times 10^2\\
L_\alpha&2.921&0.7969&0.2272&6.781&-1.171\\
H_{II}&3.167&0.7936&0.2368&0.476&-2.372\\
bcc_{II}&3.400&0.8027&0.2204&1.094&-4.038
\end{array}
$
\bigskip
\begin{description}
\item[Table II] Neutral lipid: contributions to the free energy per unit
volume and temperature in the $L_{\alpha}$ phase and in the $H_{II}$
phase, the difference in these contributions, and the derivative of this
difference with respect to the solvent fugacity.  All contributions are
evaluated at the $L_{\alpha}$, $H_{II}$ transition occurring on the
water-poor side of the azeotrope. The solvent chemical potential at the
transition is $z_s\approx 3.15$.  
All contributions are measured
with respect to the free energy of the disordered phase, ($D$).  The
temperature, $T/T_0=0.67$. ${\cal E}\equiv v_hE/VkT$,
${\cal S}_l\equiv v_hS_l/Vk$, and  ${\cal 
S}_s\equiv v_hS_s/Vk+\phi_s\ln\ z_s$.
\end{description}
\bigskip
$
\begin{array}{lrrrr}
 &L_{\alpha}-D&H_{II}-D&L_{\alpha}-H_{II}&d(L_{\alpha}-H_{II})/dz_s\\
{\cal E}&-0.6923&-0.6086&-0.0837&0.0244\\
-{\cal S}_l&1.1367&0.9123&0.2250&-0.0413\\
- {\cal S}_s&-0.6666&-0.5253&-0.1413&-0.0394
\end{array}
$
\bigskip
\begin{description}
\item[Table III] Charged lipid with both Coulomb and charge-solvent
interactions: contributions to the free energy per unit volume and
temperature in the $L_{\alpha}$ phase and in the $H_{II}$ phase, the
difference in these contributions, and the derivative of this difference
with respect to the counter ion fugacity.  All contributions are
evaluated at the $L_{\alpha}$, $H_{II}$ transition, $z_c=0.1228$, 
and are measured
with respect to the free energy of the disordered phase, ($D$). The
temperature, $T^*=0.04$, $z_s=3$, ${\cal E}_i\equiv v_hE_i/VkT$,
 with 
$i=1,2,3$ being the hydrophilic-hydrophobic, Coulomb, and charge-neutral 
interactions respectively, ${\cal S}_l\equiv v_hS_l/Vk$, ${\cal 
S}_s\equiv v_hS_s/Vk+\phi_s\ln\ z_s$, and  ${\cal 
S}_c\equiv v_hS_c/Vk+\rho_c\ln\ z_c$, 
\end{description}
\bigskip
$
\begin{array}{lrrrr}
&L_{\alpha}-D&H_{II}-D&L_{\alpha}-H_{II}&d(L_{\alpha}-H_{II})/dz_c\\
{\cal E}_1&-0.6904&-0.6059&-0.0845&0.0152\\
{\cal E}_2&0.0037&0.0024&0.0013&0.0196\\
{\cal E}_3&-0.0378&-0.0268&-0.0110&-0.1082\\
-{\cal S}_l&1.1373&0.9110&0.2263&-0.0257\\
- {\cal S}_s&-0.6365&-0.5034&-0.1331&0.0215\\
- {\cal S}_c&0.0030&0.0020&0.0010&0.0153
\end{array}
$
\newpage
\begin{description}
\item[Figure 1] Phase diagram of the neutral lipid as a function of
  dimensionless temperature, $T^*\equiv 1/2\chi N$, 
and relative headgroup volume, 
$1/(1+\gamma_t)$. In addition to the disordered phase, $D$, there are
normal and inverted body-centered cubic phases, $bcc_{I}$ and
$bcc_{II}$, normal and inverted hexagonal phases, $H_{I}$ and $H_{II}$,
and the lamellar phase $L_{\alpha}$.
\item[Figure 2] Phase diagram of a neutral lipid with
  $1/(1+\gamma_t)=0.24$ in a solvent with $\gamma_s=0.1$ as a function of
  temperature, $T/T_0$, and fraction of solvent, $\phi_s/(\phi_s)_0$,
 with $T_0$ and $\gamma_s (\phi_s)_0$ the temperature and volume 
fraction of solvent at the
 azeotrope.
\item[Figure 3] Comparison of theoretically calculated and experimentally
 measured values of the lattice parameter $D/D_0$ of the $H_{II}$ phase 
at a temperature just above the azeotrope {\it vs.} weight fraction of water,
$\phi^w_w$.
The lattice parameter at the azeotrope is denoted $D_0$.
\item[Figure 4] Comparison of theoretically calculated and experimentally 
measured values of the lattice parameter, $D(T)/D_0$, {\it vs.} absolute
temperature $T$
 (a) for two different weight fractions of lipid, $\phi^w_l$,
and (b) along coexistence with excess water. The 
absolute temperature of the azeotrope is denoted $T_0$. 
\item[Figure 5]  Lattice parameter of the $H_{II}$ and $L_{\alpha}$ phases
along their mutual coexistence as a function of absolute temperature.
 The absolute temperature of the azeotrope is denoted $T_0$. 
\item[Figure 6] Transition temperature between $L_{\alpha}$ and $H_{II}$ phases
  for the same lipid in Figure 2, but now with a charged
  headgroup. The dimensionless strength of the Coulomb interaction is
  $\beta^*=1$.The transition temperature is plotted as a function of
  the magnitude of the average charge density on the headgroups.
  The solvent fugacity is fixed at $z_s=3$.
\item[Figure 7] Transition temperature between $L_{\alpha}$ and $H_{II}$ phases
  for the same lipid as in Figure 6, but now including the 
  interaction between charges and neutral solvent of strength
  $\lambda=0.1$. The solvent fugacity is again fixed at $z_s=3.$
\item[Figure 8] a) Volume fraction distribution in the $L_{\alpha}$ phase 
of the solvent, headgroups,
  and tails, in the above system at a counter ion  chemical potential
  of $z_c=0.1228$ corresponding to the $L_{\alpha}-H_{II}$ transition. 
  The  temperature is $T^*=0.04$, and the lattice parameter of the 
  lamellar phase is $D_L/R_g = 4.14$; b) 
charge densities arising from the headgroups, the
  counter ions, and the total charge density in the $L_{\alpha}$ phase 
  under the same conditions as in a).
\item[Figure 9] a) Volume fraction distribution in the $H_{II}$ phase 
of the solvent, headgroups,
  and tails, in the above system at a counter ion  chemical potential
  of $z_c=0.1228$ corresponding to the $L_{\alpha}-H_{II}$ transition.
  The  temperature is $T^*=0.04$, and the lattice parameter of the 
  inverted hexagonal phase is $D_H/R_g = 3.83$.  b) charge densities arising
  from the headgroups, the
  counter ions, and the total charge density in the $H_{II}$ phase under
  the same conditions as in a).
\end{description}
\newpage
\begin{figure}
\epsfxsize = 6.in
\epsfysize = 6.in
\centering
\epsfbox{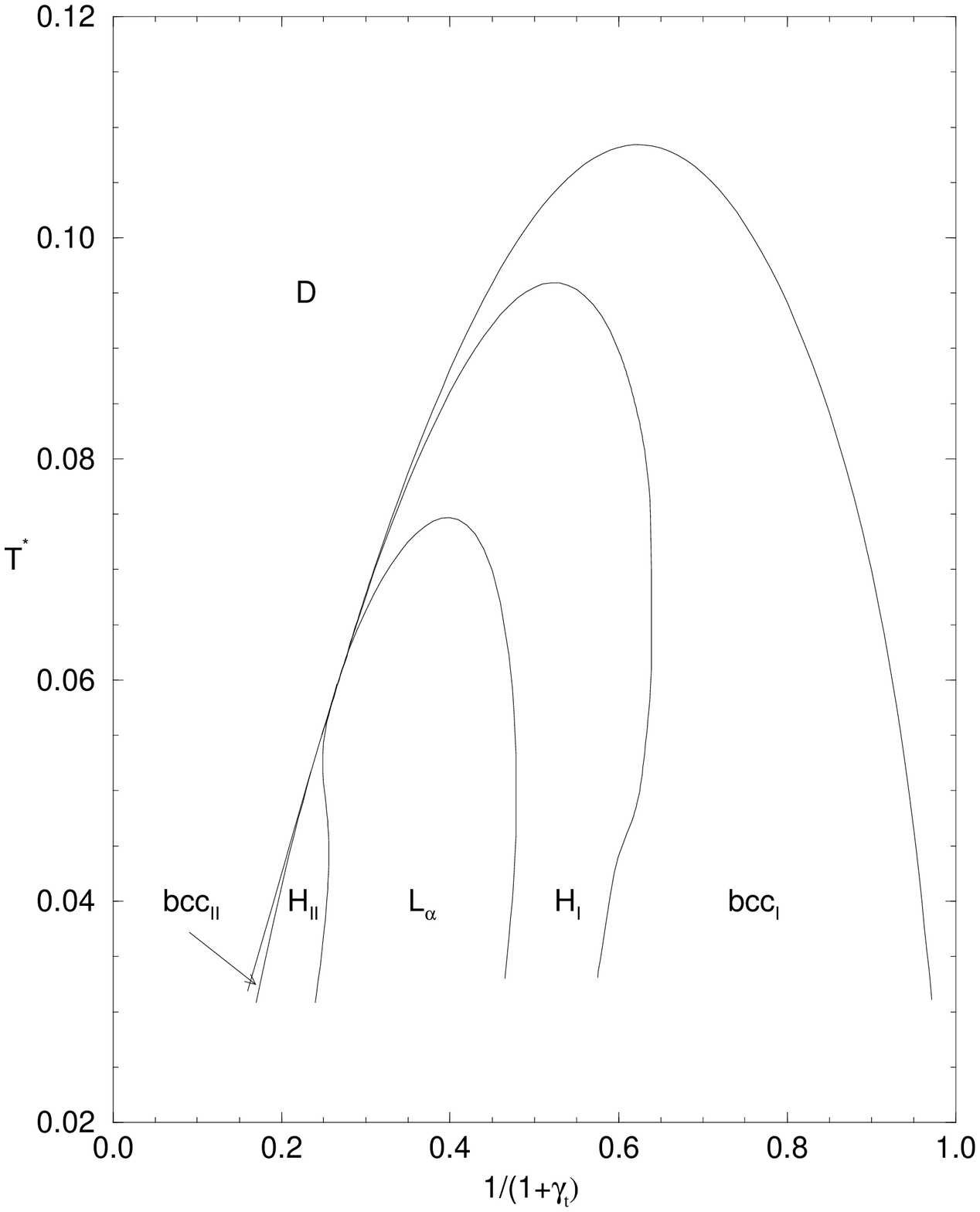}
\caption{}
\end{figure}

\newpage
\begin{figure}
\epsfxsize = 6.in
\epsfysize = 6.in
\centering
\epsfbox{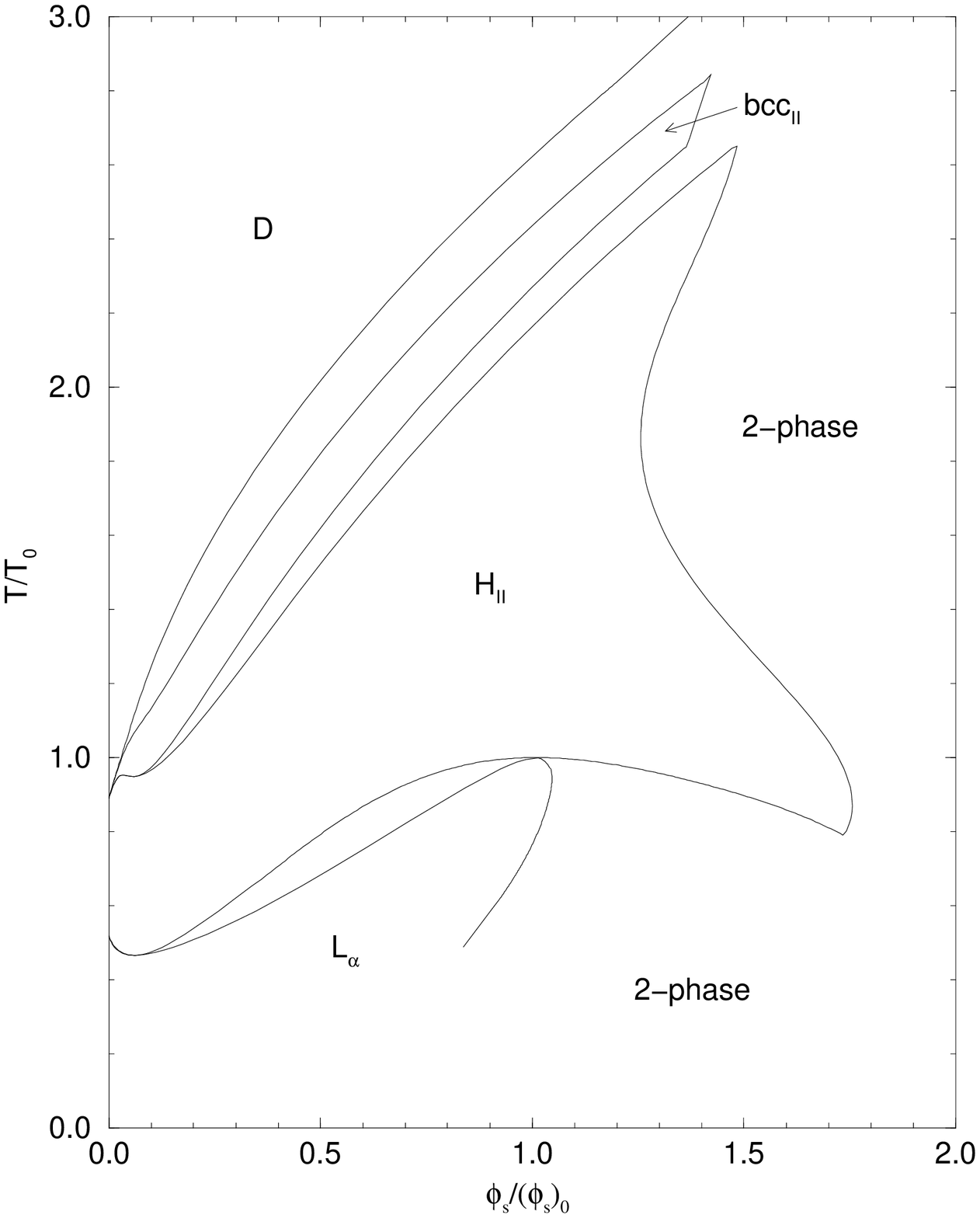}
\caption{}
\end{figure}

\newpage
\begin{figure}
\epsfxsize = 6.in
\epsfysize = 6.in
\centering
\epsfbox{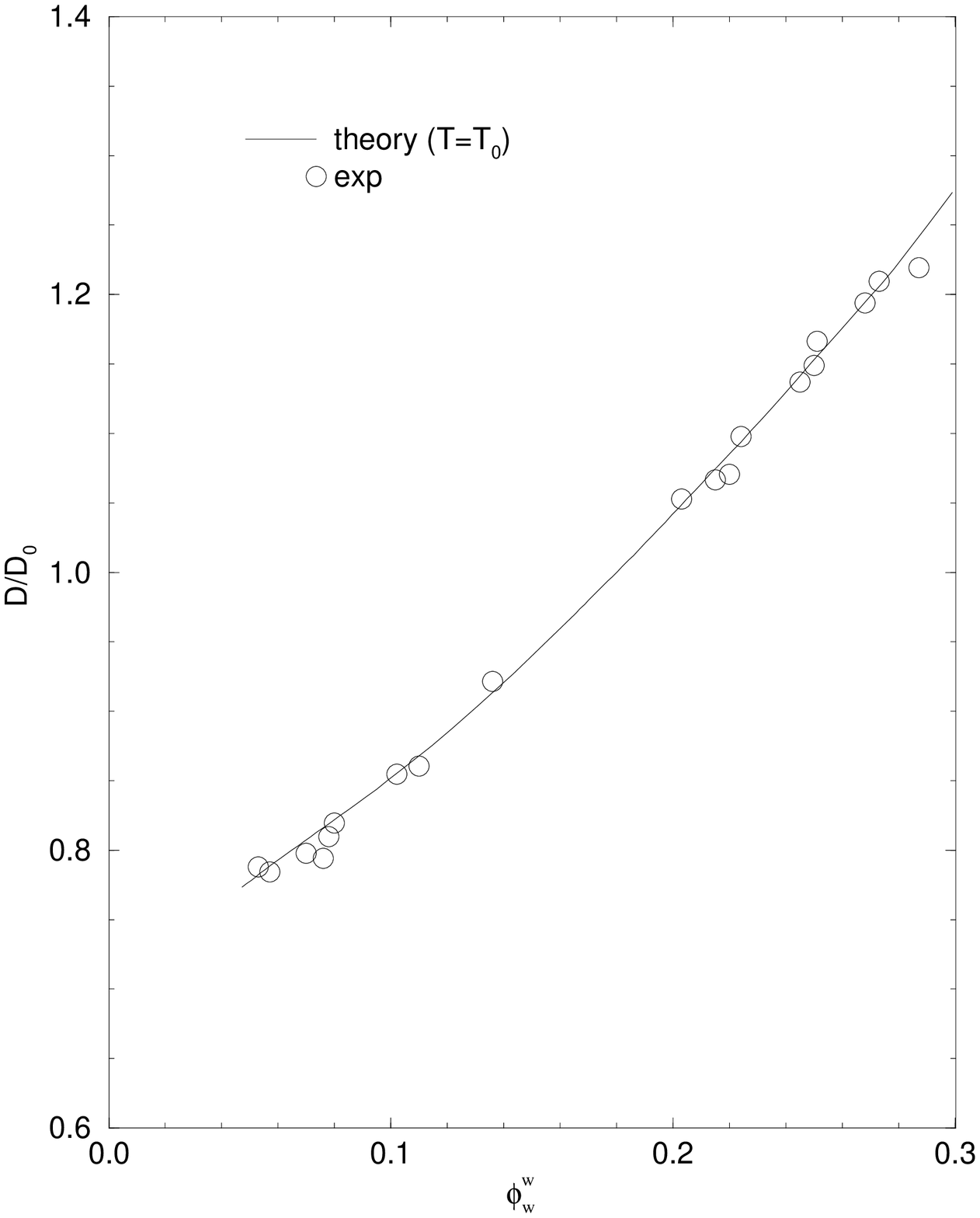}
\caption{}
\end{figure}

\newpage
\begin{figure}
\epsfxsize = 6.in
\epsfysize = 6.in
\centering
\epsfbox{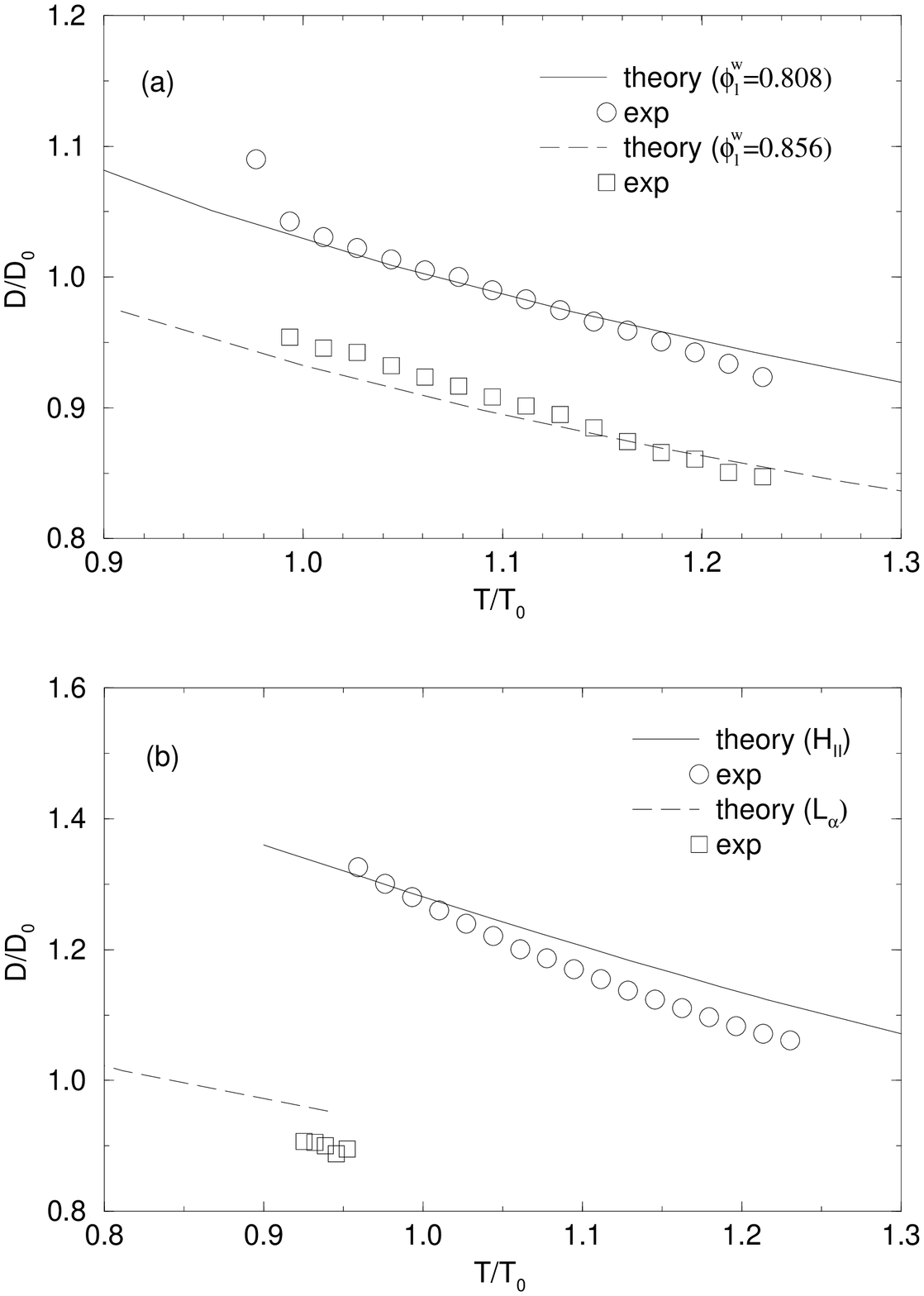}
\caption{}
\end{figure}

\newpage
\begin{figure}
\epsfxsize = 6.in
\epsfysize = 6.in
\centering
\epsfbox{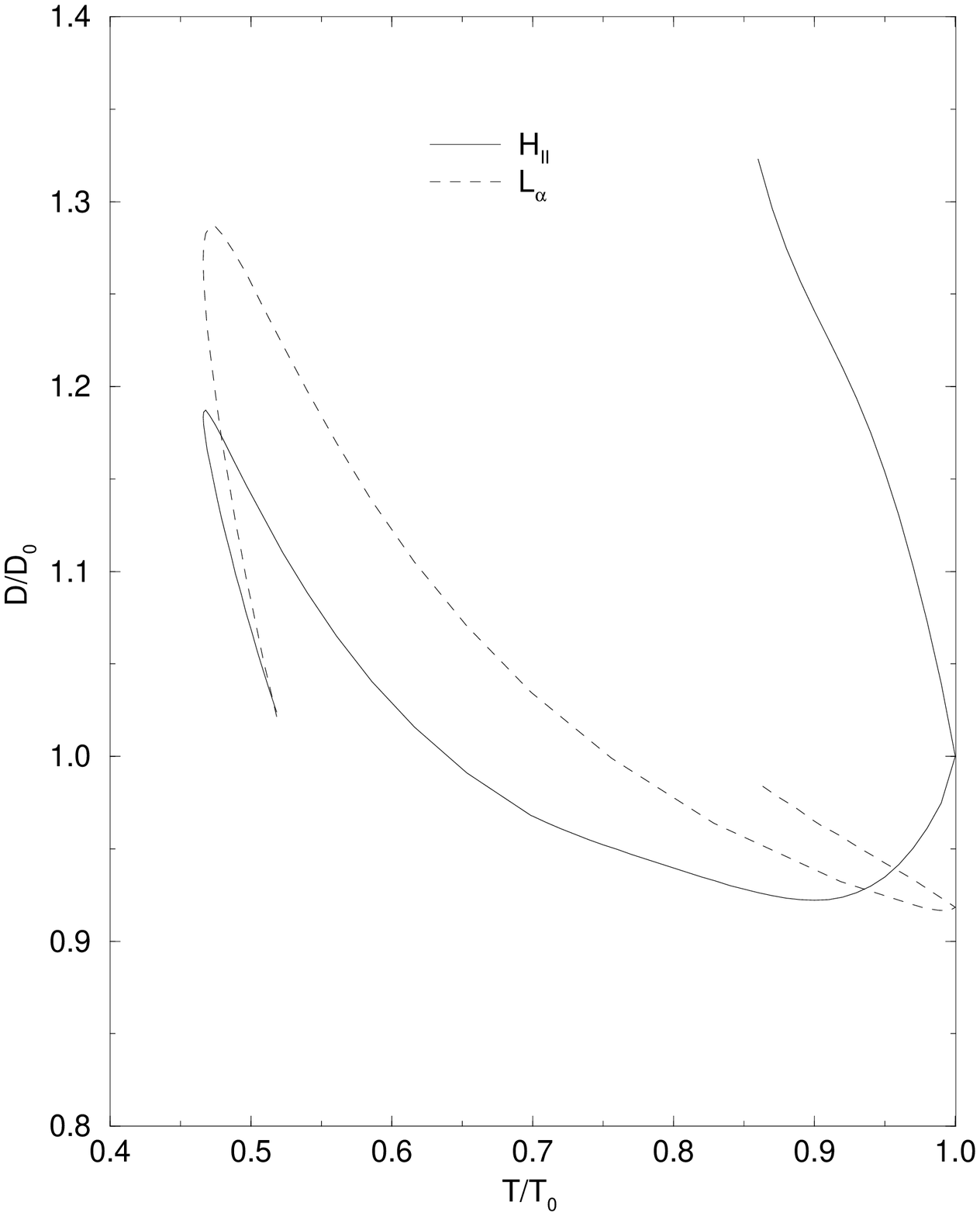}
\caption{}
\end{figure}

\newpage
\begin{figure}
\epsfxsize = 6.in
\epsfysize = 6.in
\centering
\epsfbox{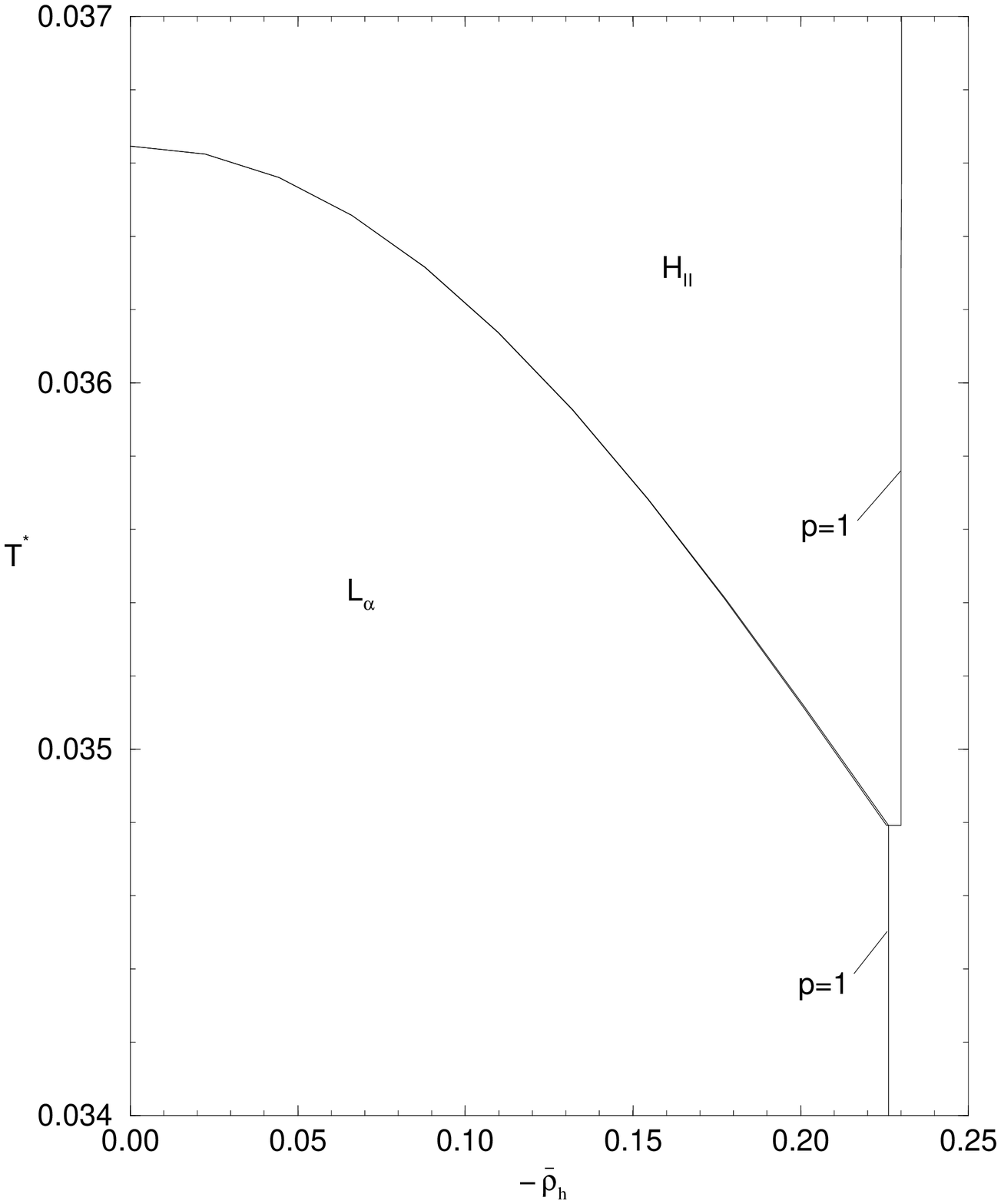}
\caption{}
\end{figure}

\newpage
\begin{figure}
\epsfxsize = 6.in
\epsfysize = 6.in
\centering
\epsfbox{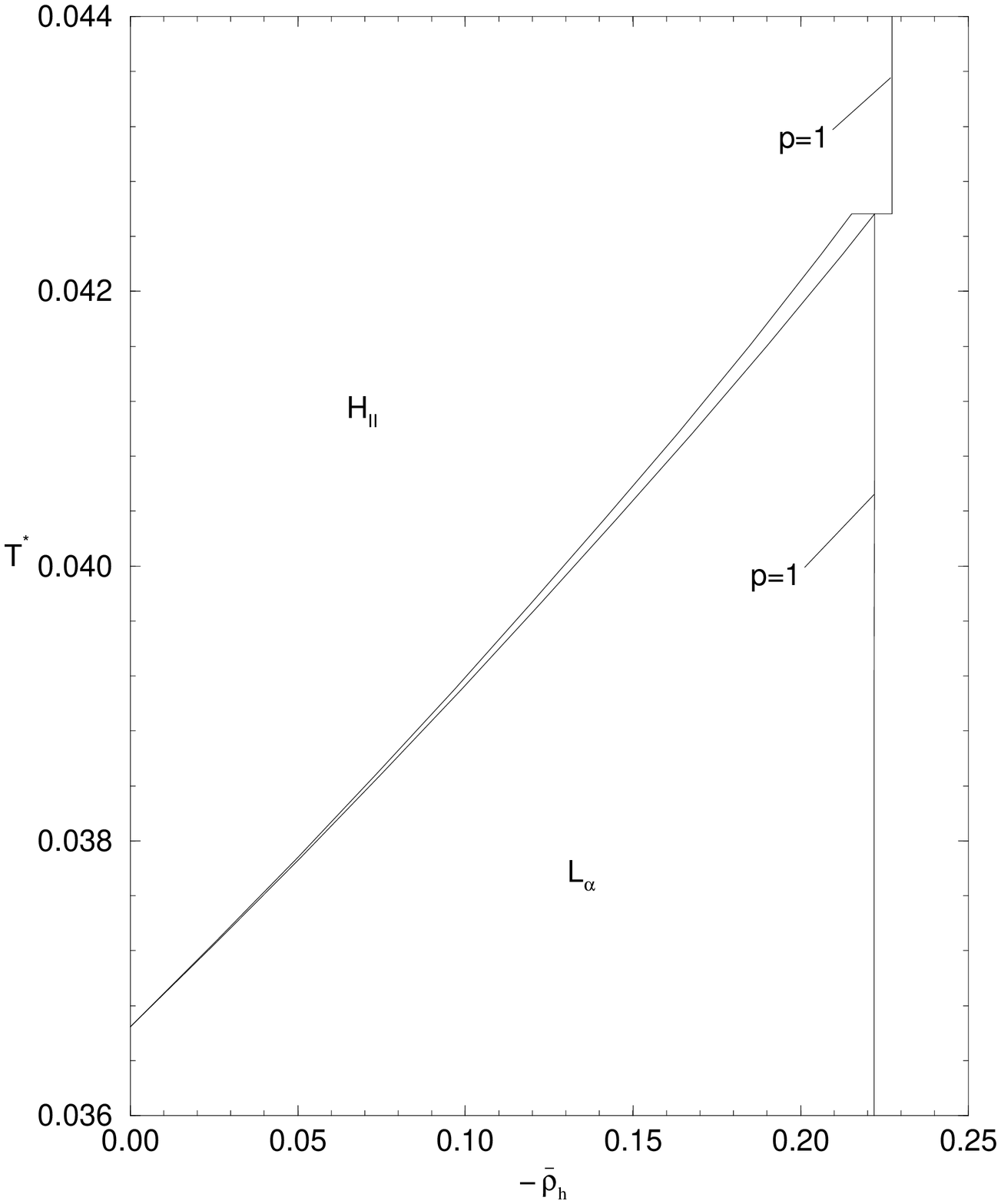}
\caption{}
\end{figure}

\newpage
\begin{figure}
\epsfxsize = 6.in
\epsfysize = 6.in
\centering
\epsfbox{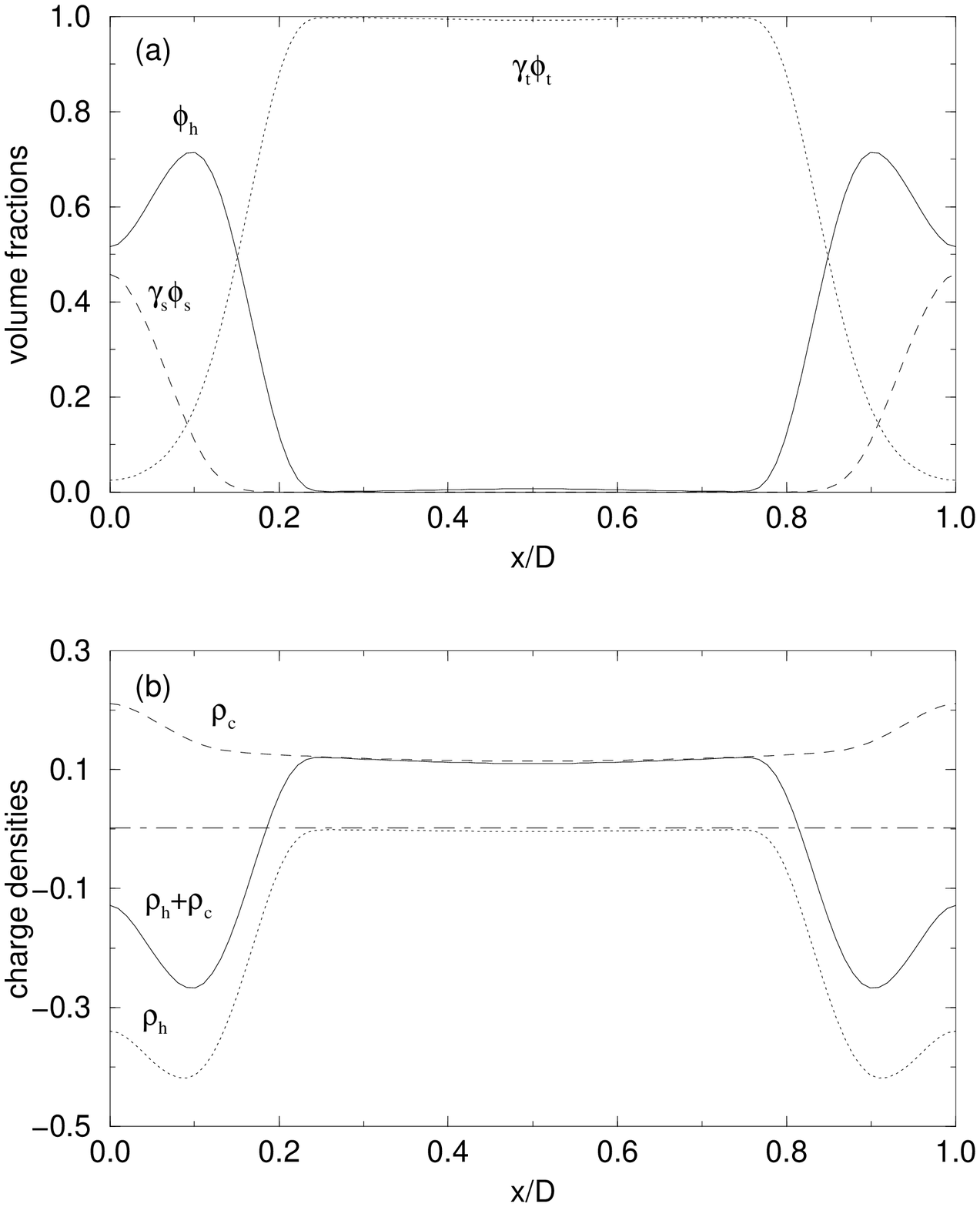}
\caption{}
\end{figure}

\newpage
\begin{figure}
\epsfxsize = 6.in
\epsfysize = 6.in
\centering
\epsfbox{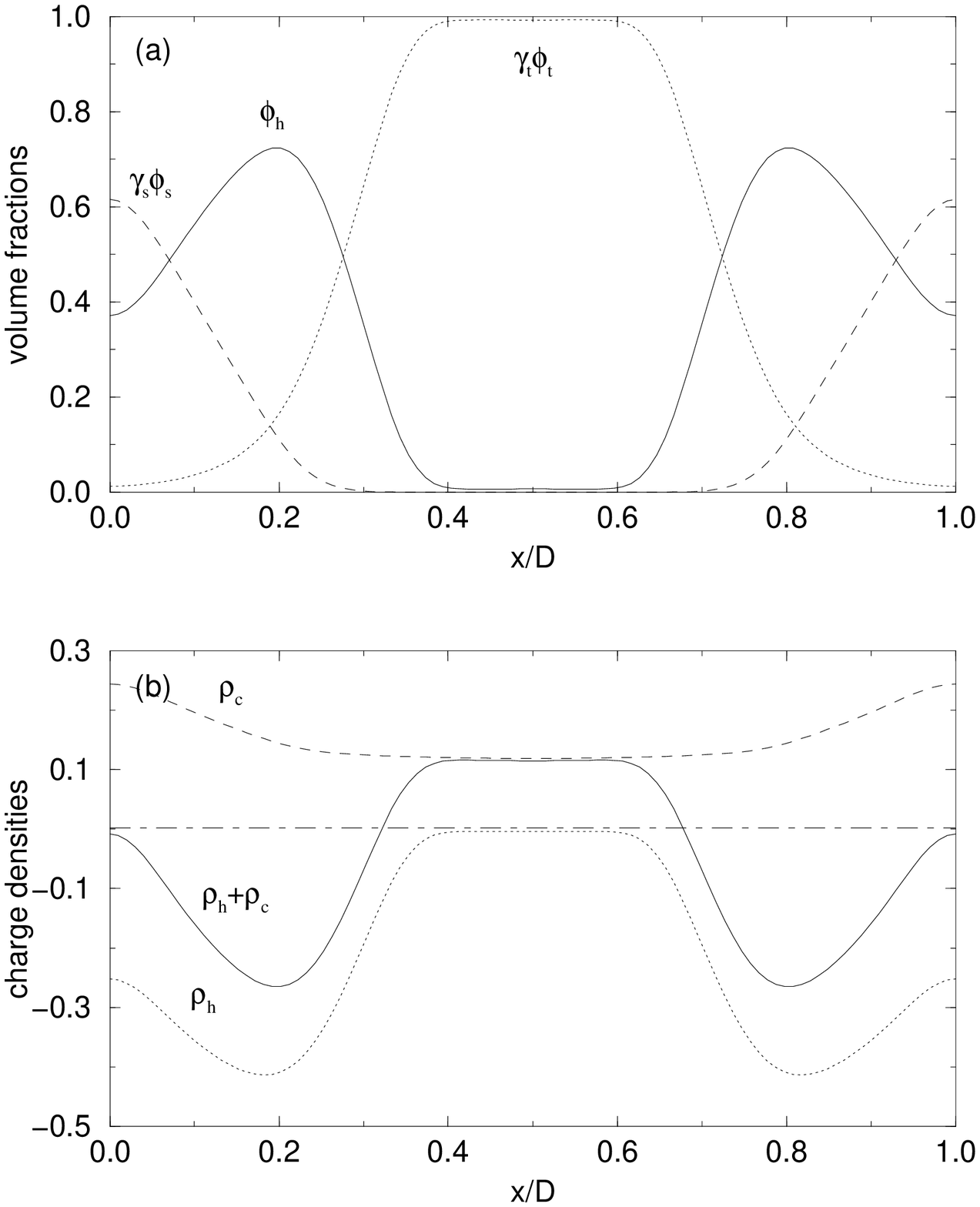}
\caption{}
\end{figure}



\end{document}